\documentclass[aps,prl,groupedaddress,onecolumn]{revtex4}

\usepackage{amsmath}
\usepackage{graphicx}
\usepackage{color}
\usepackage{gensymb}
\usepackage{datetime}
\usepackage{pgffor, pdfpages}

\begin{document}


\title{Self-sustained micromechanical oscillator with linear feedback}


\author{Changyao Chen$^{1}$, Dami\'an H. Zanette$^{2}$, Jeffrey R. Guest$^{1}$, David A. Czaplewski$^{1}$, Daniel L\'opez$^{1}$}
\affiliation{$^{1}$Center for Nanoscale Materials, Argonne National Laboratory, Lemont, USA, 60439\\
		$^{2}$Centro At\'omico Bariloche and Instituto Balseiro, Comisi\'on Nacional de Energ\'{\i}a At\'omica. Consejo Nacional de Investigaciones Cient\'{\i}ficas y T\'ecnicas. 8400 San Carlos de Bariloche, Argentina}


\pagenumbering{gobble}
\begin{abstract}
Autonomous oscillators, such as clocks and lasers, produce periodic signals \emph{without} any external frequency reference. In order to sustain stable periodic motions, there needs to be external energy supply as well as nonlinearity built into the oscillator to regulate the amplitude. Usually, nonlinearity is provided by the sustaining feedback mechanism, which also supplies energy, whereas the constituent resonator that determines the output frequency stays linear. Here we propose a new self-sustaining scheme that relies on the nonlinearity originating from the resonator itself to limit the oscillation amplitude, while the feedback remains linear. We introduce a model to describe the working principle of the self-sustained oscillations and validate it with experiments performed on a nonlinear microelectromechanical (MEMS) based oscillator.     
\end{abstract}


\maketitle

Autonomous oscillators are systems that can spontaneously commence and maintain stable periodic signals in a self-sustained manner without external frequency references. They are abundant both in Nature and in manmade devices. In Nature made systems, the self-sustained oscillators are the fundamental piece that describes systems as diverse as neurons, cardiac tissue, and predator-prey relationships \cite{Strogatz_book_2014}. In manmade devices, self-sustained autonomous oscillators are overwhelmingly used for communications, timing, computation, and sensing \cite{Jenkins_pr_2013}, with examples such as quartz watches \cite{time_book} and laser sources \cite{Maiman_nature_1960}. A typical oscillator consists of a resonating component and a sustaining feedback element: the constituent resonator determines the oscillation frequency, whereas the feedback system draws power from an external source to compensate the energy loss due to damping during each oscillation of the resonator \cite{Andronov_book_1966}. In order to initiate the oscillations, the initial gain of the feedback must be larger than unity, so that energy accumulates to build up oscillation amplitude \cite{Barkhausen_book_1963}. However, to avoid ever increasing oscillations, some limiting mechanism must act to ensure that, eventually,  the vibrational amplitude no longer grows. 

In the conventional designs of oscillators, the resonating element is operated in the linear regime, where its resonant frequency is independent of the excitation levels, and the necessary amplitude limiting mechanism is enacted in the feedback loop by introducing a nonlinear element (Fig.~1a).  However, maintaining the resonating element in the linear regime has been challenging for a variety of applications requiring self-sustained oscillators made from micro-/nano-electromechanical (M/NEMS) resonators \cite{naik_nnano_2009,Ekinci_rsi_2005,Cottone_prl_2009}, mostly because these resonators exhibit significantly reduced linear dynamic range. To limit the amplitude, common mechanisms include impulsive energy replenishment \cite{Andronov_book_1966}, saturated gain medium \cite{Townes_prl_1958,Maiman_nature_1960} or amplifiers \cite{Vittoz_book_2010,Feng_nnano_2008_private}, automatic level control \cite{Lee_ieee_2003,Lin_ieee_2004}, phase locked loops \cite{Giessibl_rmp_2003,Antonio_ncomm_2012}, nonlinear signal transduction \cite{Nguyen_jssc_1999}, and dedicated nonlinear components \cite{Villanueva_nl_2011}. These mechanisms to incorporate nonlinear elements into the electronic feedback circuitry introduce technical challenges in the analysis, design and implementation of the oscillators due to the significant impedance mismatch between CMOS drivers and M/NEMS resonators \cite{vanBeek_jmm_2011}.

In this Letter, we introduce and analyze a new oscillator architecture that solely relies on the nonlinearity originated from a micromechanical resonator, while all components of the feedback circuitry stay within the \emph{linear} regime  (Fig.~1b). By capitalizing on the intrinsic nonlinear dynamics of the mechanical resonator, it is possible to considerably simplify the design of the oscillator while achieving a large degree of control and tunability. Unlike the techniques used with linear resonators, the oscillator architecture we are proposing can be readily implemented in practically all M/NEMS geometries, as the only requirement is the existence of a nonlinear response. The proposed innovative architecture permits to (1) initiate the oscillation spontaneously, (2) achieve stable oscillations through interplay between elastic nonlinearities and viscous damping, and (3) tune the oscillation frequency over a wide range with readily accessible system parameters. We demonstrate this new architecture with an oscillator consisting of a clamped-clamped (c-c) silicon MEMS resonator \cite{Antonio_ncomm_2012} with high quality factor ($Q \gtrsim 10^5$), and with frequency tunability as large as 19 \%. 

We treat the mechanical resonator as a generic single degree-of-freedom oscillating element, whose departure from equilibrium  is described by a coordinate $x(t)$ obeying \cite{Lifshitz_2009}:
\begin{equation}
m\ddot{x} + (\gamma + \tilde{\eta} x^2)\dot{x} + m \omega_0^2 x + \tilde{\beta} x^3 = F(x, \dot{x}),
\label{eq_EOM}
\end{equation}
where $m$ is the effective mass, $\gamma$ and $\tilde{\eta}$ are the linear and nonlinear damping
coefficients \cite{Dykman_pss_1975,Dykman_book_2012}, $\omega_0$ is the natural frequency of linear oscillation, $\tilde{\beta}$ is the cubic (Duffing) nonlinear coefficient, and $F(x, \dot{x})$ is the driving force from feedback.
Since we only focus on periodic solutions,  quadratic nonlinearities are ignored. To facilitate  the analysis, we define $\epsilon  q^{-1} = \gamma/m\omega_0$, $\eta= \tilde{\eta}/4\gamma$, $\beta = 3\tilde{\beta}/4m\omega_0^2$, and a fast time scale $\tau = \omega_0 t$. Here, the small expansion parameter $\epsilon$ is introduced for treatment within a perturbation theory, as shown below. Since the feedback force, $F$, is only needed to compensate the dissipation, it will also be of the order of $\epsilon$. Furthermore, we treat the feedback force as proportional to the vibrational amplitude, corresponding to the cases where the vibrational amplitude is linearly transduced and directly measured experimentally. Similarly, the feedback force can be treated as proportional to the vibrational velocity, if the velocity is the observed quantity as in the case of capacitive motion transduction \cite{prl_SI}. For simplicity, here we consider the case of linear amplitude amplification and scale the feedback force as $F = \epsilon m\omega_0^2 g x \cos{\Delta}$, where $g$ is the feedback gain and $\Delta$ the feedback loop phase-delay. With these definitions,  Eq. (\ref{eq_EOM})  becomes
\begin{equation}
\ddot{x} +\epsilon q^{-1}(1+4\eta x^2) \dot{x} + x + \frac{4}{3}\beta x^3 = \epsilon g x \cos{\Delta}.
\label{eq_EOM2}
\end{equation}
Here the time derivatives are calculated with respect to $\tau$.

We represent the limit of small dissipation by taking $\epsilon \ll 1$, and the scaled quality factor $q$ of the order of unity. 
The solution to Eq.~(\ref{eq_EOM2}) can be found through perturbation theory \cite{Nayfeh_book_2008}. However, in contrast with previous treatments with weak nonlinearity \cite{Lifshitz_2009}, we do not assume that the cubic force is small as compared to the linear term \cite{Arroyo_epjb_2016}. The resulting zeroth-order equation is therefore the nonlinear Duffing equation without damping:
 $\ddot{x}_0 + x_0  + \frac{4}{3} \beta x_0^3= 0$. We propose a steady-state solution of the form $x_0 = A_0 \cos \Omega_0 \tau$, where $A_0$ and $\Omega_0$ are the oscillation amplitude and frequency of the zeroth-order solution, respectively. By neglecting higher-harmonics contributions, we find that $A_0$ and $\Omega_0$ must satisfy the relation
\begin{equation} \label{eq_backbone}
1 - \Omega_0^2 +\beta A_0^2 = 0.
\end{equation}

\begin{figure}
 \includegraphics[width = 0.5\textwidth]{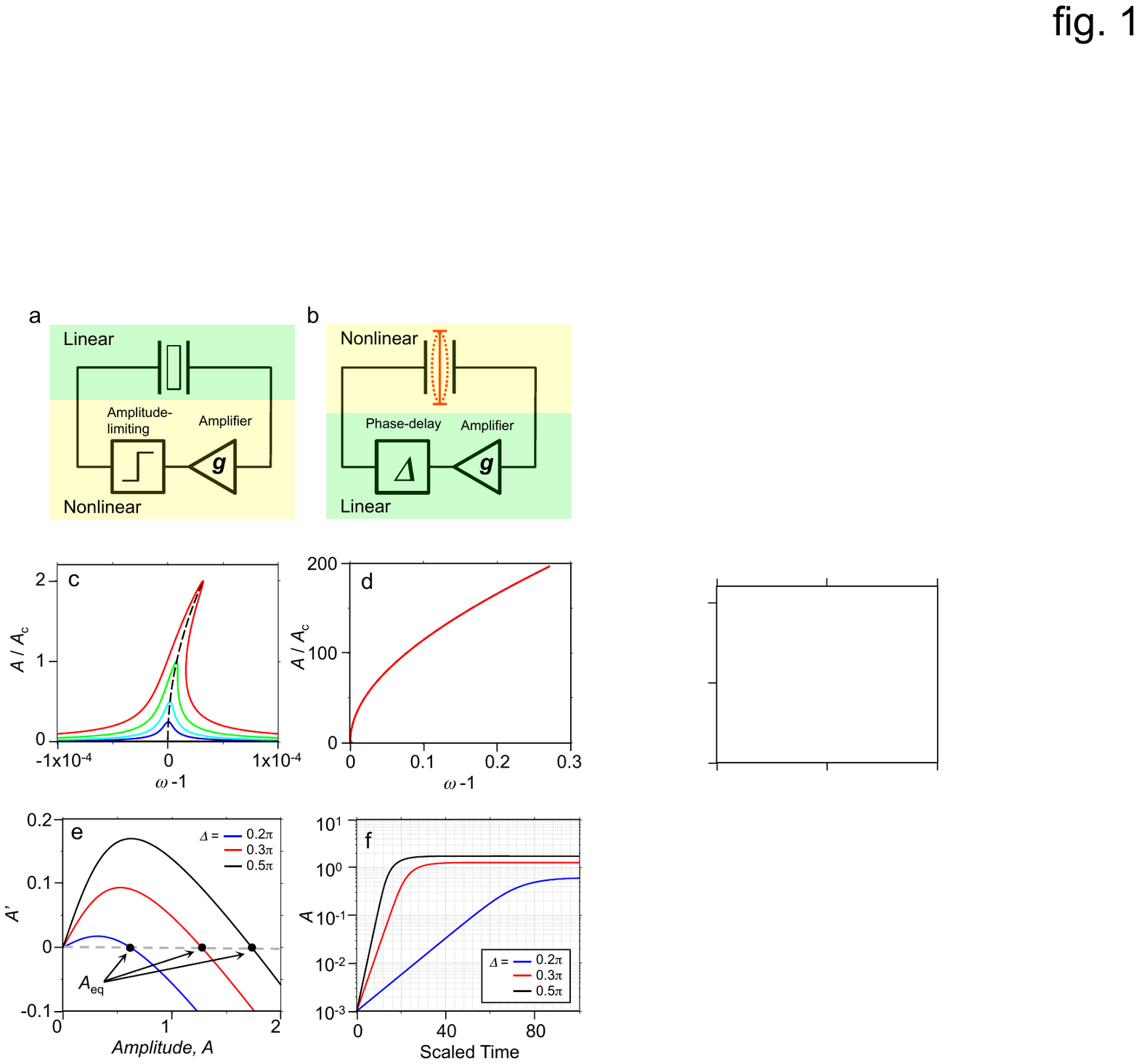}
 \caption{(a) Schematics of the conventional oscillator design, consisting of a linear resonator and nonlinear feedback. The output of the resonator is first amplified, and then amplitude-limited before being fed-back to the resonator. (b) Schematics of oscillator design with a nonlinear resonator and linear feedback loop. The output of the resonator is amplified and phase shifted, and then re-injected to the resonator. (c) Vibrational amplitude $A$ (expressed as the ratio to the critical amplitude $A_c$, above which multiple solutions exist) versus frequency detuning in the limit of small drive, derived from Eq.~(\ref{eq_EOM2}), with $\epsilon$ = 10$^{-5}$, $\beta$ = 1, $\eta$ = 0, and different levels of excitation. The dashed backbone curve shows the solution to Eq.~(\ref{eq_backbone}). (d) Same results but with large excitation, where the full solution practically coincides with the backbone. (e) Phase portrait of amplitude $A$, showing unstable rest states, and different stable equilibrium $A_\text{eq}$ at various $\Delta$. (f) Simulated transient responses of $A$ at different $\Delta$.   \label{model}}
\end{figure}

Following the method of multiple time-scales \cite{Nayfeh_book_2008, prl_SI}, we proceed to obtain the steady-state solution to Eq.~(\ref{eq_EOM2}) up to $O(\epsilon)$. This solution will be characterized by amplitude $A$ and frequency $\omega$, determined both by intrinsic properties ($\beta, \eta, q$) and extrinsic parameters $\Delta, g$ \cite{prl_SI}. Fig.~1c and 1d illustrate the relation between the steady-state amplitude $A$ and frequency $\omega$, for different values of feedback excitations and $\Delta$. For small excitations (Fig.~1c), we recover the resonance curve as obtained for weakly nonlinear resonators  \cite{Lifshitz_2009}. For large excitation (Fig.~1d), where the oscillation frequency is pulled far away from the linear resonance ($\omega - 1 \gg \epsilon$), the perturbed solution is practically identical to that of zeroth-order equation Eq.~(\ref{eq_backbone}). It is worth noting that, in the case of our autonomous oscillator, both the amplitude $A$ and the frequency $\omega$ are functions of the phase delay $\Delta$. Therefore, if multiple solutions of $A$ exist for a given value of $\omega$, all of them are stable \cite{Villanueva_prl_2013,prl_SI}, as opposed to an externally driven resonator where only two solutions are stable. Moreover, nonlinearities make it possible to achieve frequencies far above the  linear oscillation frequency, only bounded by other nonlinearities present in the system, or by physical limits of the device. 

For hardening nonlinearity ($\beta > 0$), closer inspection of the full solution reveals that the zero-amplitude state for our system is unstable, as indicated by the phase portrait shown in Fig.~1e: any disturbance will push the oscillator away from rest state, towards the stable equilibrium $A_\text{eq}$, marked by the arrows in Fig.~1e. Physically, this stable equilibrium can be understood as the energy balance between driving and damping: if the amplitude $A$ increases suddenly around $A_\text{eq}$, (for instance, by noise), due to the proportionality between the forcing and the amplitude there is a growth in the input energy. At the same time, because of  the hardening nonlinearity, the increased amplitude pushes the oscillation frequency $\omega$ upward, resulting in a larger energy dissipation due to viscous damping, which is proportional to the product of $A\omega$. The energy balance is thus restored, and the oscillation amplitude achieves a steady state (Fig.~1f). A symmetric argument holds if the amplitude decreases. We have also examined the case where the feedback force is proportional to velocity, and arrived at similar conclusion \cite{prl_SI}. In this case,  however, the balance is maintained by the effects of nonlinear damping, which becomes stronger as the oscillation amplitude grows. This energy balance highlights the key difference with previous oscillator topologies with a dedicated amplitude limiting element \cite{Feng_nnano_2008_private,Villanueva_nl_2011,Antonio_ncomm_2012}. In the current setup, intrinsic nonlinearities are an indispensable ingredient for stabilization. Additionally, the spontaneous oscillation greatly simplifies the startup protocol of the oscillator, making it highly suitable for M/NEMS based oscillators where very sensitive transducers are required to initiate the motion.

We used a MEMS based oscillator to experimentally demonstrate these concepts. The resonator, similar to the one used in \cite{Antonio_ncomm_2012}, is placed in a vacuum chamber and actuated electrostatically. The mechanical vibration creates a capacitive current in the sensing comb, that is proportional to the velocity. Both of the comb electrodes consist of 25 interdigitated fingers that allow efficient excitation and linear signal transduction. The measured linear resonance is 61.57 kHz, with linear damping rate of 0.51 Hz \cite{prl_SI}. The small dissipation of $\epsilon^{-1} = Q \sim$ 120,000 ensures that the resonator is well suited for the designed nonlinear oscillator. In this case, the nonlinearity is geometrical in origin and arises from the elongation of the beam during large transverse vibration. The onset of nonlinearity $x_c$ - above which the amplitude-frequency relation bifurcates - is calculated to be 17 nm from the geometry of the device \cite{Ekinci_rsi_2005}, and experimentally found to be about 10 nm \cite{prl_SI}. In the experiments, we have observed oscillation amplitudes larger than 1 $\mu$m, well above the linear threshold. We find excitations larger than 100 $\mu$V are enough to drive the resonator into the nonlinear regime. When the resonator is excited with an even larger force, it displays the signature of nonlinear damping \cite{Lifshitz_2009,Imboden_pr_2014,Polunin_jmems_2016,prl_SI}.

 \begin{figure}
 \includegraphics[width = 0.5\textwidth]{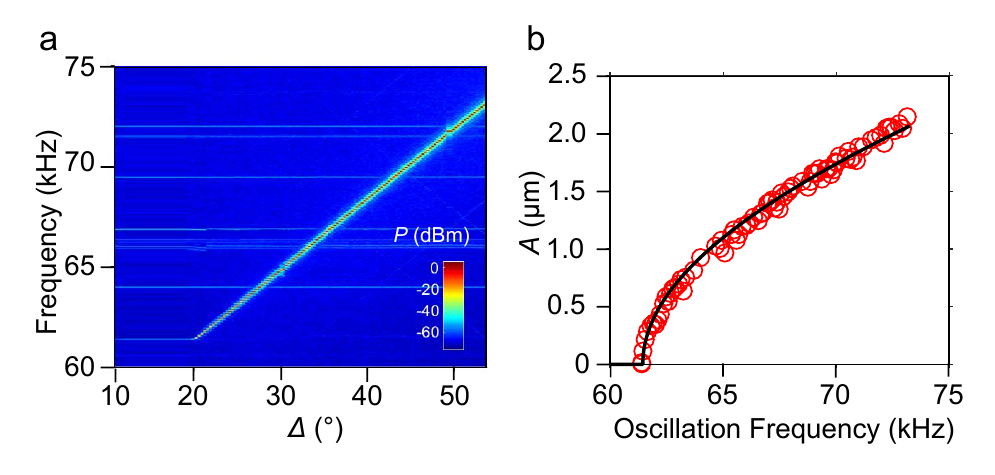}
 \caption{Steady-state response of a nonlinear oscillator. (a) Measured power spectrum of the oscillation with different feedback-phase-delay $\Delta$. Measurement at each $\Delta$ is performed with a time constant long enough to ensure the transient response has died out. The DC bias is 7 V. (b) Extracted steady-state oscillation amplitude and frequency (red circles), and fitting to Eq.~(\ref{eq_backbone}), with $\beta = 1.15\times 10^{11} $m$^{-2}$.  \label{fig_ss}}
 \end{figure}
 
The feedback loop consists of a transimpedance amplifier followed by a voltage amplifier and a band pass filter. Therefore, the feedback force is proportional to the velocity with a certain phase-delay. In order to ensure the linearity of the feedback loop, we have calibrated the linearity of each component in the feedback circuity, and found all of them operating in the linear regime \cite{prl_SI}. Operating the electronics in the linear regime provides a large range of operational voltages that allows for significant detuning of the MEMS. Additionally, eliminating complex controlling circuits for oscillators \cite{Sedra_book_1998} greatly reduces the number of elements in the feedback loop, lowering the power consumption \cite{Nguyen_ieee_2007}.

Figure \ref{fig_ss}a shows the steady-state power spectrum of the oscillation, measured at different phase-delay $\Delta$. For $\Delta < $ 20$^{\circ}$, no  oscillation is observed, whereas for $\Delta > $ 20$^{\circ}$, the oscillations occur and the frequency grows monotonically with $\Delta$. This is consistent with the fact that, in order to initiate spontaneous oscillations, the feedback force should overcome damping. The onset of the oscillation frequency is about 61.5 kHz, slightly above the linear resonance, and the highest oscillation frequency observed is 73.15 kHz, which is about 19 \% above the linear resonance. We are hindered by the instrumental limit from achieving larger phase-delay and frequency detuning. The acquisition time of each spectrum is much longer than the transient time of the oscillation, to ensure steady-state conditions. The oscillation amplitude versus frequency (Fig.~\ref{fig_ss}b) clearly shows the quasi-square-root dependency, as predicted by Eq.~(\ref{eq_backbone}). The scaled Duffing nonlinearity $\beta$, obtained from fitting to Eq.~(\ref{eq_backbone}), is $1.15\times 10^{11} $m$^{-2}$, in good agreement with previous result \cite{Antonio_prl_2015}. 

 \begin{figure}
 \includegraphics[width = 0.46\textwidth]{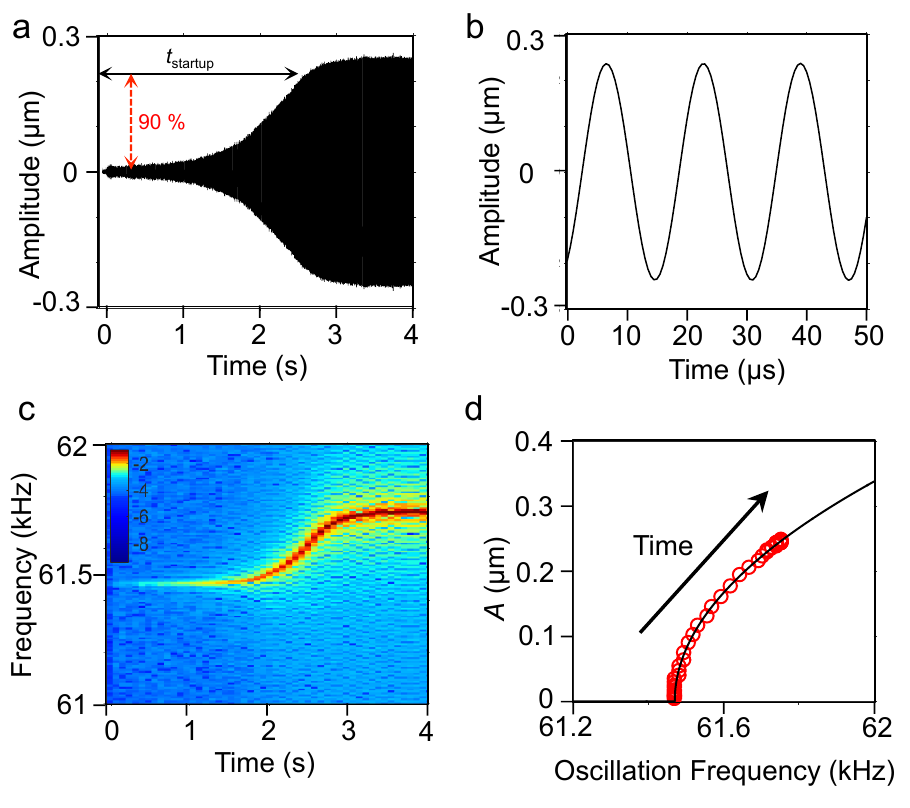}
 \caption{Transient response during startup. (a) Measured amplitude during oscillation buildup. The steady-state oscillation frequency is 61.7 kHz, with $\Delta$ = 24.1$^{\circ}$. (b) Zoomed-in view of the steady-state oscillation of (a). The $x$-axis is shifted arbitrarily.  (c) Temporal frequency response of the oscillation. The power spectrum at each nominal time $t_i$ is obtained by performing non-overlapping fast Fourier transform (FFT) of the time domain data in a narrow window around $t_i$. (d) The temporal evolution of the oscillator on the amplitude-frequency plane, and fitting to Eq.~(\ref{eq_backbone}) (black solid line). The extracted $\beta$ is $1.05\times 10^{11} $m$^{-2}$. The DC bias is 7 V for all the data shown. \label{fig_ringup}}
 \end{figure}
 
Next we consider the buildup of the oscillation. The spontaneous initiation of the oscillating motion with linear feedback does not require the prerequisite of the Barkhausen criterion \cite{Villanueva_nl_2011}: after the amplified and phase-shifted signal is fed back to actuate the resonator, the system will asymptotically transition to the stable limit cycle, whose frequency can be controlled by $g$ and $\Delta$ \cite{prl_SI}. Fig.~\ref{fig_ringup} shows the temporal evolution of the oscillator during the startup: after the feedback is engaged at $t$ = 0 s, the envelope of oscillation amplitude, $A$, grows rapidly towards the final value (Fig.~\ref{fig_ringup}a). The steady-state response shows stable sinusoidal oscillation, as shown in Fig.~\ref{fig_ringup}b. The temporal frequency evolution shows a similar pattern (Fig.~\ref{fig_ringup}c, corresponding to the time domain data shown in Fig.~\ref{fig_ringup}a): the instantaneous frequency starts at the linear resonant frequency value, and shifts upward towards the steady-state oscillation frequency. This temporal evolution is shown on the amplitude-frequency plane (Fig.~\ref{fig_ringup}d), where we plot the amplitude-frequency of the oscillator every 80 miliseconds. It can be clearly seen that the temporal response of the oscillator follows the prescribed square-root interdependence, Eq. (3), with an extracted $\beta = 1.05\times 10^{11} $m$^{-2}$. The inter-dependence between the vibrational amplitude and frequency underlines the working principle of the stable oscillation: any unintentional increase in amplitude will increase the resonant frequency of the resonator, which leads to more viscous damping, which in turn reduces the amplitude, hence maintaining the oscillations stable.

 \begin{figure}[t]
 \includegraphics[width = 0.5\textwidth]{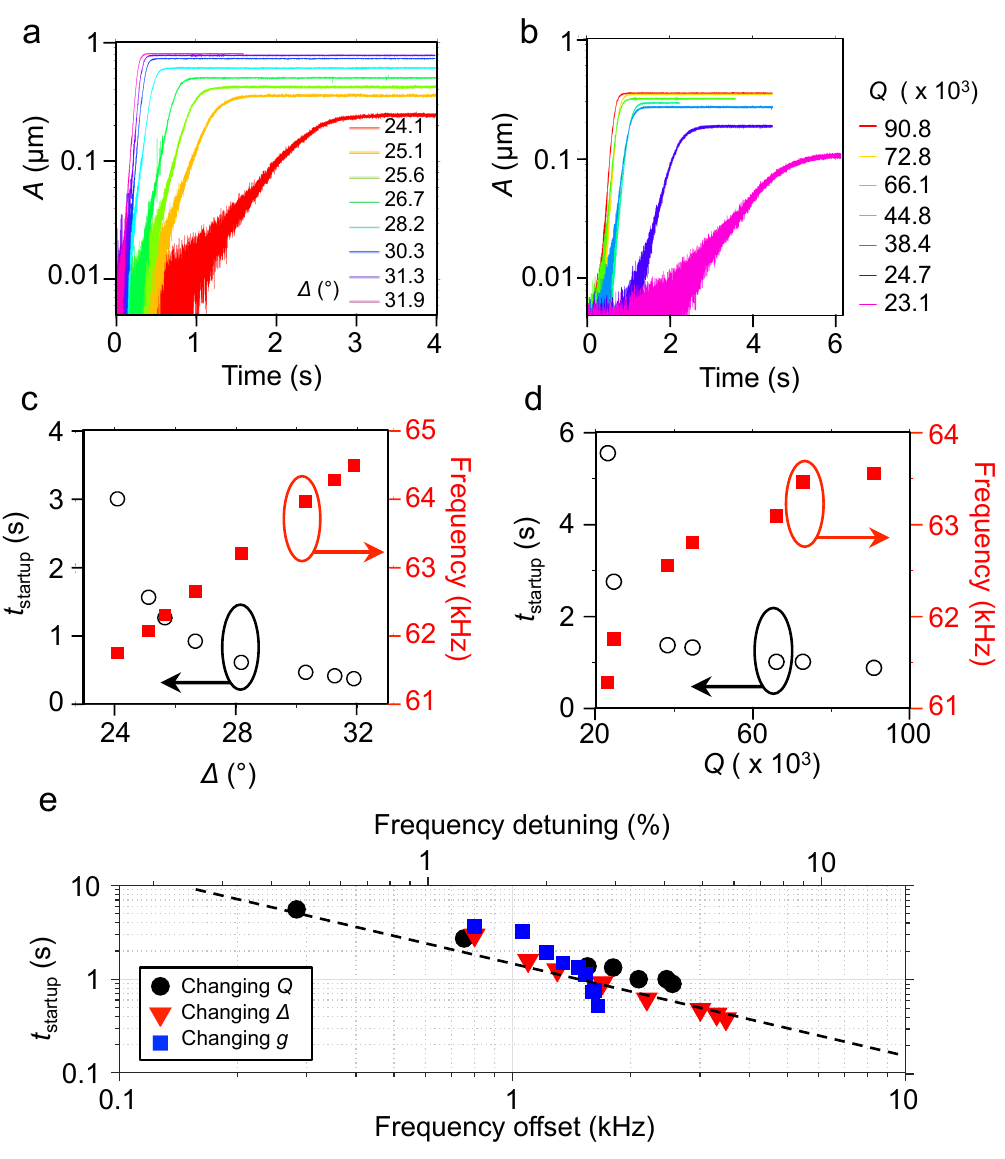}
 \caption{Control of oscillation startup. (a) Envelope of the amplitude during startup, for different $\Delta$, with $Q$ = 120,000. The envelope is obtained through averaging of multiple cycles of the oscillation around given times. (b) Envelope of the amplitude during startup, for different $Q$, with $\Delta$ = 27$^{\circ}$. (c, d) Corresponding startup time $t_\text{startup}$ and steady-state frequency extracted from (a) and (b), respectively. (e) The startup time $t_\text{startup}$ versus steady state frequency offset, collected with different startup conditions. The dashed line is a guide to the eyes with slope of $-1$.}
 \end{figure}

Finally, we will examine the influence of the system parameters on the oscillation buildup. We define the startup time, $t_\text{startup}$, as the time needed for the oscillation amplitude to reach 90\% of its steady-state value. We find that $t_\text{startup}$ drops considerably with increasing the phase-delay $\Delta$ (Fig.~4a, c), which is consistent with the theoretical modeling considering the feedback force is proportional to $g$ and ${\Delta}$ \cite{prl_SI}. Additionally, we also modify the effective gain $g$ by changing the DC bias and observe a similar dependence of $t_\text{startup}$ \cite{prl_SI}.

To  validate the model against intrinsic properties of the resonator, we deliberately tune the linear damping rate $\gamma$ by increasing the pressure of the vacuum chamber, which changes the quality factor $Q$. The values of $Q$ are obtained from separated open-loop resonator-type measurements \cite{prl_SI}. The startup time increases drastically when $Q$ drops below $\sim$ 30,000 (Fig.~4b, d),  and we failed to observe any oscillation for $Q < 10,000$. 

Since the only requirement for this oscillator topology to work is to have the resonator in the nonlinear regime, self-sustained oscillations can be achieved at low values of $Q$ by changing the dimensions of the resonator. The onset of nonlinearity scales with the characteristic length of the resonating element, and for NEMS devices, even forces from thermal noise can drive the resonator into the nonlinear regime \cite{Gieseler_nphys_2013}. 

The startup time, obtained from different system configurations, is shown in Fig.~4e, and is found to be approximately inversely proportional to the steady state frequency offset. This observation highlights another benefit of the new topology: the more nonlinear the response is, the shorter the startup time. The nonlinear mechanical resonator ensures stable oscillation, whose frequency offset is proportional to the total feedback gain, whereas the linear feedback allows for exponentially fast transient towards the stable oscillation, which results in shorter startup time with larger gain.
 
In summary, we have introduced a novel oscillator architecture consisting of a nonlinear mechanical resonator driven by a linear feedback loop.  We have theoretically examined the conditions for stable periodic motion and have shown that when the feedback forcing is proportional to the vibration amplitude, a hardening nonlinear response ensures the balance between external energy input and intrinsic dissipation necessary to stabilize oscillations. The interplay between the resonator's frequency dependent amplitude and the associated damping underlines the principle of stable oscillations. When the feedback forcing is proportional to the oscillation velocity, in turn, the balance is guaranteed by nonlinear damping. 

As the size of the resonating elements shrinks towards the nanoscale, the critical amplitudes for onset of nonlinearity decreases accordingly  \cite{Ekinci_rsi_2005,Postma_apl_2005} and the resonators will inevitably operate in the mechanical nonlinear regime, even merely driven by thermal noise\cite{Gieseler_nphys_2013}. Since the mechanical nonlinearities of the resonator are responsible for achieving self-sustained oscillations, the new architecture should perform better when scaled down to the nanoscale \cite{Chen_nnano_2013}, making it ideal for oscillators incorporating nanoscale resonators and for very large-scale integration of high-$Q$ MEMS and NEMS. 

Use of the Center for Nanoscale Materials at the Argonne National Laboratory was supported by the U.S. Department of Energy, Office of Science, Office of Basic Energy Sciences, under Contract No. DE-AC02-06CH11357. We thank D. Antonio for helpful discussions.

\nocite{Kovacic_book_2011, Lifshitz_2009, Villanueva_nl_2011, Polunin_jmems_2016, Imboden_nl_2013, Imboden_pr_2014, Antonio_ncomm_2012}



\section*{Supplementary material}

[36] M. Imboden, O. A. Williams, and P. Mohanty, Nano letters **13**, 4014 (2013).

# Supplementary Information for
## *A self-sustained micro mechanical oscillator with linear feedback*

Changyao Chen, Damián H. Zanette, Jeffrey R. Guest, David A. Czaplewski, Daniel López

## 1 Equilibrium and stability analysis

Eq. (1) in the main text considers the force balance in a mechanical oscillator. On the left hand side, it describes the intrinsic properties of the resonator, which includes the inertial force ($\ddot{x}$ term), friction ($\dot{x}$ term), and elastic restoring force ($x$ and $x^3$ terms). The $x^3$ term, also known as the Duffing term, is the consequence of treating the system as a perturbed simple harmonic oscillator with a first order correction[1]. On the right hand side, it describes the external force that is needed to maintained the self-sustained motion, and for an autonomous system, this feedback force is an implicit function of time. Since the external force is only needed to compensate the energy loss ($\dot{x}$ term on the left hand side), it will be of the same order of the friction. In real mechanical systems, the specific values of the coefficients in Eq. (1) can differ by orders of magnitude (see Table S1 below). In order to discuss different MEMS and NEMS in a generic manner, it is advantageous to scale the coefficients in Eq. (1) with corresponding masses and resonant frequencies, to arrive at Eq. (2) in the main text.

Using multiple time scale method, we propose a solution of $x(\tau, T) = A(\tau, T)\cos[\phi(\tau, T)] = A(\tau, T)\cos[\Omega(\tau, T)\tau + \alpha(\tau, T)]$ to the scaled governing equation of motion (Eq. (2) in main text),

$$\ddot{x} + \epsilon q^{-1}(1 + 4\eta x^2)\dot{x} + x + \frac{4}{3}\beta x^3 = \epsilon f_0 \cos[\phi + \Delta], \quad (S1)$$

where the dot indicates time derivative with respect to $\tau$, and $T = \epsilon\tau$ is the slow time scale. Therefore $\frac{d}{d\tau} = \epsilon\frac{d}{dT}$. Note that here we express the feedback force with its amplitude, $f_0$, and its phase difference with respect to the motion, $\Delta$, in order to facilitate later analysis. Using secular

Table S1. Comparison of system parameters in different MEMS/NEMS

| Material | $m$ (kg) | $\gamma$ (kg/s) | $Q$ | $\tilde{\eta}$ (kg/m²s) | $\omega_0$ (rad/s) | $\tilde{\beta}$ (kg/m²s²) | Reference |
|---|---|---|---|---|---|---|---|
| CNT | 7.87 x 10$^{-21}$ | 6.30 x 10$^{-16}$ | 20000 | 1.00 x 10$^4$ | 1.60 x 10$^9$ | 6.00 x 10$^{12}$ | [a] |
| Graphene | 3.90 x 10$^{-19}$ | 8.70 x 10$^{-14}$ | 5630 | 1.50 x 10$^7$ | 1.26 x 10$^9$ | 1.40 x 10$^{16}$ | [a] |
| Diamond | 1.23 x 10$^{-14}$ | 4.59 x 10$^{-10}$ | 3520 | 5.00 x 10$^8$ | 1.31 x 10$^8$ | 3.36 x 10$^{11}$ | [b] |
| AlN | 2.74 x 10$^{-16}$ | 2.37 x 10$^{-11}$ | 6750 | n/a | 5.83 x 10$^8$ | 3.75 x 10$^{14}$ | [c] |
| SiC | 7.22 x 10$^{-16}$ | 4.98 x 10$^{-12}$ | 8000 | n/a | 5.51 x 10$^7$ | 7.33 x 10$^{13}$ | [d] |
| Pt | 4.64 x 10$^{-17}$ | 2.19 x 10$^{-12}$ | 6045 | n/a | 2.85 x 10$^8$ | 3.34 x 10$^{13}$ | [e] |
| SiN | 8.60 x 10$^{-14}$ | 3.52 x 10$^{-11}$ | 13600 | 8.60 x 10$^2$ | 5.56 x 10$^6$ | 1.72 x 10$^{10}$ | [f] |
| Silicon | 2.06 x 10$^{-10}$ | 8.73 x 10$^{-10}$ | 90900 | 2.33 x 10$^2$ | 3.85 x 10$^5$ | 6.03 x 10$^{12}$ | This work |

[a] Eichler, A., *et al*. "Nonlinear damping in mechanical resonators made from carbon nanotubes and graphene." *Nature Nanotechnology* 6.6 (2011): 339-342.
[b] Imboden, Matthias, Oliver A. Williams, and Pritiraj Mohanty. "Observation of nonlinear dissipation in piezoresistive diamond nanomechanical resonators by heterodyne down-mixing." *Nano Letters* 13.9 (2013): 4014-4019.
[c] Aldridge, J. S., and A. N. Cleland. "Noise-enabled precision measurements of a Duffing nanomechanical resonator." *Physical Review Letters* 94.15 (2005): 156403.
[d] Kozinsky, I., *et al*. "Tuning nonlinearity, dynamic range, and frequency of nanomechanical resonators." *Applied Physics Letters* 88.25 (2006): 253101.
[e] Kozinsky, I., *et al*. "Basins of attraction of a nonlinear nanomechanical resonator." *Physical Review Letters* 99.20 (2007): 207201.
[f] Zaitsev, Stav, *et al*. "Nonlinear damping in a micromechanical oscillator." *Nonlinear Dynamics* 67.1 (2012): 859-883.

perturbation theory, the $O(1)$ equation is

$$1 - \Omega^2 + \beta A^2 = 0, \tag{S2}$$

which coincides with the "backbone" of the nonlinear resonator². Eq. (S2) is also rendered as Eq. (3) in the main text, with subscript of 0. If $\beta$ is small and considered on the order of $\epsilon$, then the term $\beta A^2$ will be absent from Eq. (S2), and the $O(1)$ equation corresponds to a simple harmonic oscillator with solution of $\Omega = 1$.

We next look at the $O(\epsilon)$ equations. By arranging coefficients of $\sin\phi$ and $\cos\phi$, we have

$$2(A'\Omega + A\Omega') + q^{-1}A\Omega(1 + \eta A^2) = f_0 \sin\Delta, \tag{S3}$$

$$2A\Omega(\Omega'\tau + \alpha') = -f_0 \cos\Delta, \tag{S4}$$

where the prime represents the time derivative with respect to $T$.

The equilibrium is reached at $A' = 0$ which, according to Eq. (S2), also leads to $\Omega' = 0$. Then, the oscillation amplitude at equilibrium $A_{\text{eq}}$ can be obtained by solving

$$q^{-1} A_{\text{eq}} (1 + \eta A_{\text{eq}}^2) \sqrt{1 + \beta A_{\text{eq}}^2} = f_0 \sin \Delta, \tag{S5}$$

and the apparent frequency $\omega = \dot{\phi} = \Omega + \epsilon(\Omega' \tau + \alpha')$ can subsequently be calculated as

$$\omega = \sqrt{1 + \beta A_{\text{eq}}^2} - \frac{\epsilon f_0 \cos \Delta}{2 A_{\text{eq}} \sqrt{1 + \beta A_{\text{eq}}^2}}. \tag{S6}$$

We can further eliminate $\Delta$ from Eqs. (S5) and (S6) to obtain the relation between the oscillation frequency and amplitude. For $\Delta = \pi/2$, it corresponds to the maximum of $A$, as well as of the velocity $U \ (\equiv A\Omega)$.

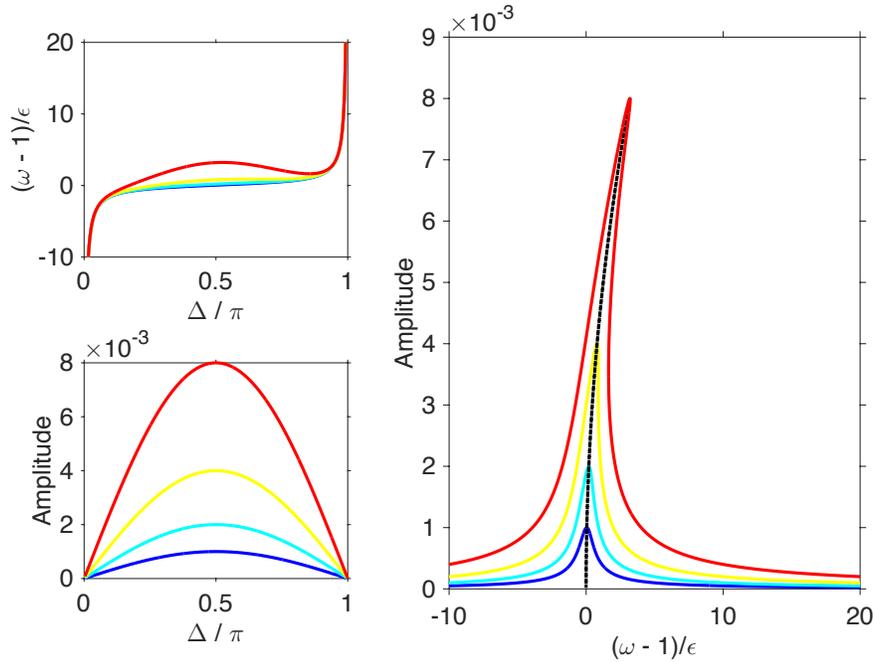

**Figure S1** Parametric plots of apparent frequency $\omega$, equilibrium amplitude $A$ as functions of phase delay

3

$\Delta$. Here $\omega$ is expressed in the unit of linewidths away from linear resonance. The forcing are 0.001, 0.002, 0.004 and 0.008, $\epsilon = 10^{-5}$, $q = 1$, $\beta = 1$, and $\eta = 0$.

To determine whether $A'_{eq} = 0$ is a stable equilibrium, we note that $\Omega' = \beta A'A/\Omega$ and obtain:

$$A' = \frac{f_0 \sin \Delta - q^{-1} A\Omega(1+\eta A^2)}{2(\Omega + \beta A^2/\Omega)}. \tag{S7}$$

In the following, we will discuss three different scenarios, where the forcing amplitude $f_0$ is 1) independently controlled, 2) proportional to the amplitude, and 3) proportional to the velocity.

$f_0$ **is independently controlled**. This applies to the condition, for example, of an external phase locked loops (PLL) circuit being used to set the forcing. In such case,

$$\frac{dA'}{dA} = -\frac{1+\eta A^2}{2q} - \frac{\eta A^2 \Omega^2}{(2\Omega^2 - 1)q} - \frac{A\beta(2\Omega^2 + 1)[f_0 \sin \Delta - q^{-1} A\Omega(1+\eta A^2)]}{2\Omega(2\Omega^2 - 1)^2}. \tag{S8}$$

Since at equilibrium the last term in Eq. (S8) vanishes, Eq. (S8) becomes

$$\left.\frac{dA'}{dA}\right|_{A=A_{eq}} = -\frac{1+\eta A^2}{2q} - \frac{\eta A^2 \Omega^2}{q(2\Omega^2 - 1)} \tag{S9}$$

Therefore, this equilibrium is globally stable, even for $\eta = 0$. Here we ignore the case of extremely nonlinear softening where $\Omega^2 < 1/2$. The presence of nonlinear damping will improve stability.

$f_0$ **is proportional to the vibrational amplitude**. We next consider that the forcing is applied as linear amplification of the vibration amplitude, such that $f_0 = gA$, where $g$ is the gain. This scheme applies if the vibration amplitude can be directly measured, for example, through piezoresistive transduction[3]. Invoking Eq. (S7), we have

$$\frac{dA'}{dA} = -\frac{\Omega^2 - 1}{2q(2\Omega^2 - 1)} - \frac{\eta A^2}{2q} + \frac{g \sin \Delta - q^{-1}\Omega(1+\eta A^2)}{2\Omega(2\Omega^2 - 1)^2}. \tag{S10}$$

At equilibrium, the last term in Eq. (S10) vanishes, and

$$\left.\frac{dA'}{dA}\right|_{A=A_{eq}} = -\frac{\Omega^2 - 1}{2q(2\Omega^2 - 1)} - \frac{\eta A^2}{2q} \tag{S11}$$

4

Thus for hardening nonlinearity ($\beta > 0$), which implies $\Omega > 1$, the equilibrium state is stable with $\eta = 0$. Again, $\eta > 0$ will enhance the stability. In contrast to the case where $f_0$ is independently controlled, here we have an equilibrium ($A' = 0$) at $A = 0$, with $dA'/dA = \frac{1}{2}(g \sin \Delta - q^{-1})$. Therefore, if the feedback gain is large enough to overcome linear damping, such that $g \sin \Delta - q^{-1} > 0$, the oscillation can be spontaneously initiated from rest.

**$f_0$ is proportional to the vibrational velocity**. Finally, we examine the case where the feedback force is proportional to the velocity, as in our experiment. In this case, $f_0$ can be written as $gA\Omega$, where $g$ is again the gain. From Eq. (S7) we derive

$$\frac{dA'}{dA} = -\frac{\eta A^2 \Omega^2}{q(2\Omega^2 - 1)} + \frac{[g \sin \Delta - q^{-1}(1 + \eta A^2)](2\Omega^4 - 3\Omega^2 + 2)}{2(2\Omega^2 - 1)^2}. \tag{S12}$$

Again, at equilibrium, the last term is zero. However, to have stable oscillation, we need $\eta > 0$, namely, stability is *only possible* with nonlinear damping. For the startup condition from $A = 0$, $dA'/dA = \frac{1}{2}(g \sin \Delta - q^{-1})$, which is identical to the previous case.

## 2   Characteristics of nonlinear damping

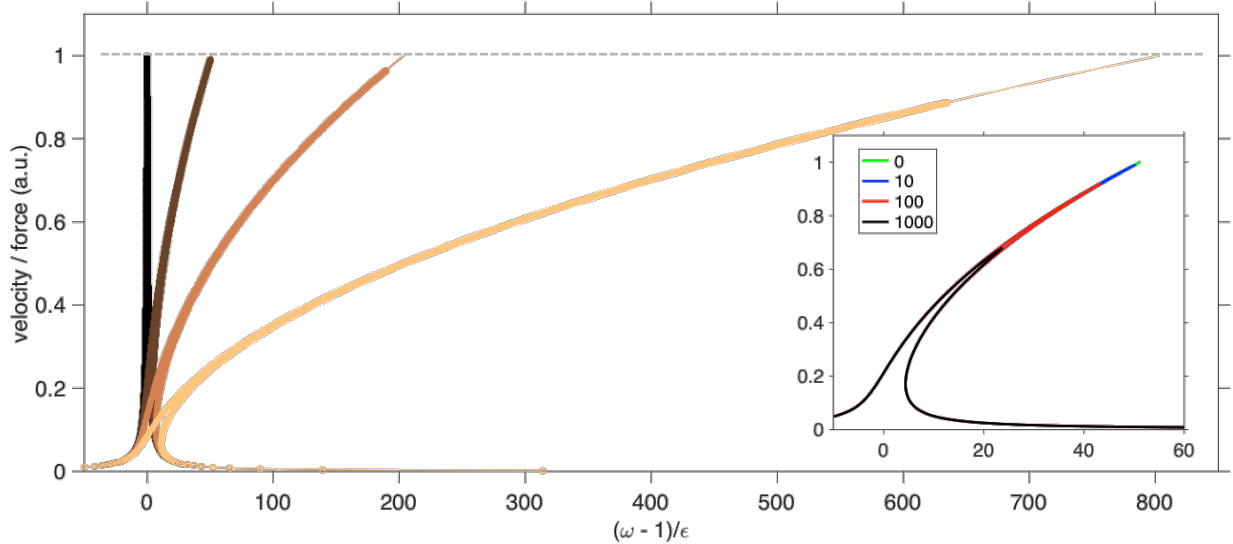

**Figure S2**   Effects of nonlinear damping, seen as the ratio between velocity and force, at different oscillation frequencies. The thin lines are solutions to Eq.s (S3) and (S4) with $\eta = 0$, while the circles are

5

results for $\eta$ = 10. The applied forces $f_0$ are 0.001, 0.032, 0.064, 0.128, respectively. Inset: same result for different values of $\eta$ (shown in legend), with $f_0$ = 0.032. Other parameters are: $\epsilon = 10^{-5}$, $\beta = 1$.

Since the linear velocity feedback scheme only works when nonlinear damping is present, it is important to identify its presence (see section 5 for experimental evidence). One method is to examine the equilibrium response of the resonator through quasi-static frequency sweep (open-loop), and inspect the velocity gain (ratio between velocity and applied force). Without nonlinear damping, the maximum velocity gain should always be equal to one[2], corresponding to $\Delta = \pi/2$, whereas with nonlinear damping, the maximum gain will be less than one, as shown in fig. S2. We can use such characteristic as the evidence of nonlinear damping.

Additionally, nonlinear damping will affect the dynamical response of the oscillator during ring-down[4]. When the external forcing is removed after the system has reached stable oscillation ($f_0$ in Eq. (S3) is switched to zero), there is no longer energy influx to compensate dissipation. Hence, the vibrational amplitude $A$ is expected to decay, resembling the ring-down phenomena observed in linear resonators. From Eq. (S3), it is found that ring-down is described by $2(A\Omega)' = -q^{-1}A\Omega(1 + \eta A^2)$. For $\eta = 0$. The velocity $U$ ($\equiv A\Omega$) decays exponentially, where $A$ and $\Omega$ are still inter-locked to each other as governed by Eq. (S2). In the meantime, the apparent frequency $\omega$ is reduced to $\Omega$. Therefore, the state of the oscillation will stay on the backbone during ring-down, such that both the amplitude and frequency decay almost exponentially. This is in sharp contrast with the linear case, where only the amplitude decays exponentially, with decay rate of $1/2q$, and the frequency stays constant. For $\eta > 0$, there is no closed-form solution, except for small $A$ where $\Omega \approx 1$. Under such approximation, the transient response of velocity $U(T)$ can be written as

$$U(T) = \frac{U(0)e^{-T/2q}}{\sqrt{1 + \eta U(0)^2(1 - e^{-T/q})}}, \tag{S13}$$

indicating a faster-than-exponential decay during the initial transient.

6

## 3  Dynamical behavior of the nonlinear oscillator

The ring-down behavior is independent of the specific feedback schemes, but rather sensitive to nonlinear damping, as discussed in the preceding section. By setting $f_0$ to zero, we are able to numerically evaluate the transient responses of velocity, amplitude, and frequency during the ring down, as shown in fig. S3.

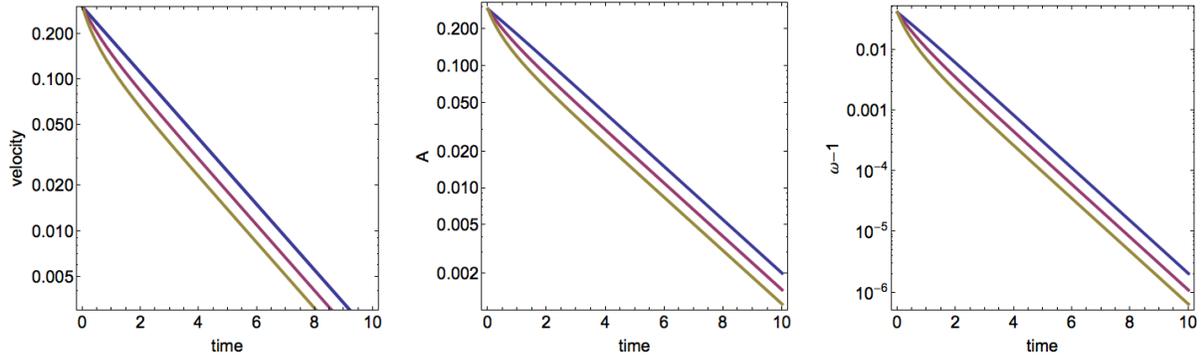

**Figure S3**  Transient responses of velocity, amplitude, and frequency offset during ringdown, for system with $\eta = 0$ (light blue), 10 (magenta) and 25 (yellow). The parameters are $\epsilon = 10^{-5}$, $\beta = q = 1$, and the initial velocity is 0.3.

For both linear feedback schemes, the dynamics of the startup are governed by the same equations. For the linear amplitude feedback scheme, given the gain is large enough to compensate the energy loss ($g \sin \Delta - q^{-1} > 0$) and the nonzero initial conditions (small value of $A$ due to noise), the oscillation will spontaneously begin, before settling to the stable equilibrium dictated by Eq. (S3) and (S4). We have numerically simulated such start-up behavior, with and without nonlinear damping, as shown in fig. S4.

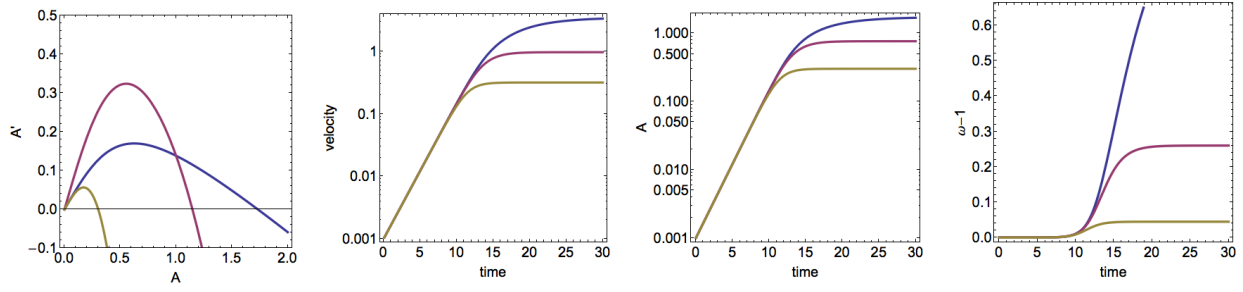

**Figure S4** Transient response during initiation for linear amplitude feedback. Parameters are: $\epsilon = 10^{-5}$, $\beta = q = 1$, $g = 2$, $\Delta = \pi/2$, $\eta = 0$ (blue), 1 (magenta) and 10 (yellow).

The start-up dynamics of the velocity feedback scheme, which is only permitted with nonlinear damping, is also examined with numerical simulations, as shown in fig. S5, with different $\Delta$. The transient responses of the velocity and frequency successfully capture the experimental observations (fig. 4 in main text).

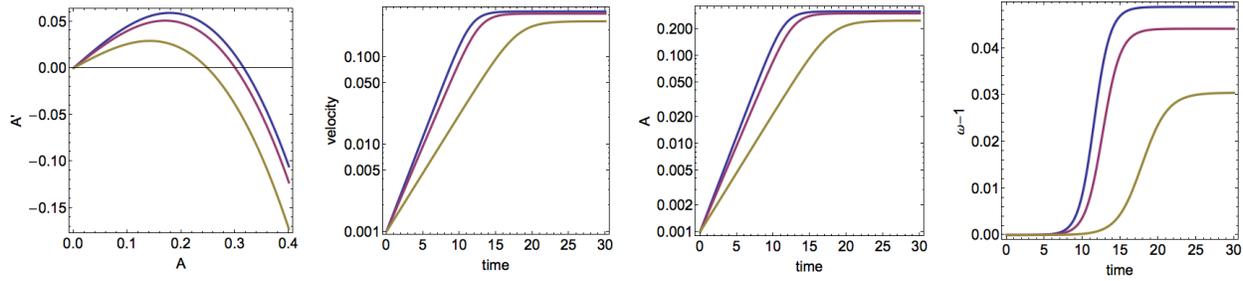

**Figure S5** Transient response during initiation. Parameters are: $\epsilon = 10^{-5}$, $\beta = q = 1$, $g = 2$, $\eta = 10$, $\Delta = 0.5\pi$, (blue), $0.4\pi$ (magenta) and $0.3\pi$ (yellow).

## 4 Linear and weakly nonlinear response of the resonator

Figure S6 shows the measured resonator response with small ac excitations, showing the transition from the linear to the nonlinear regime. The extracted onset of nonlinearity is about 10 nm, which agrees well with the calculated value of 17 nm. The smaller observed value can be attributed to the loss of signal in the transmission. The data is taken with a Zurich Instruments HF2LI lock-in amplifier.

8

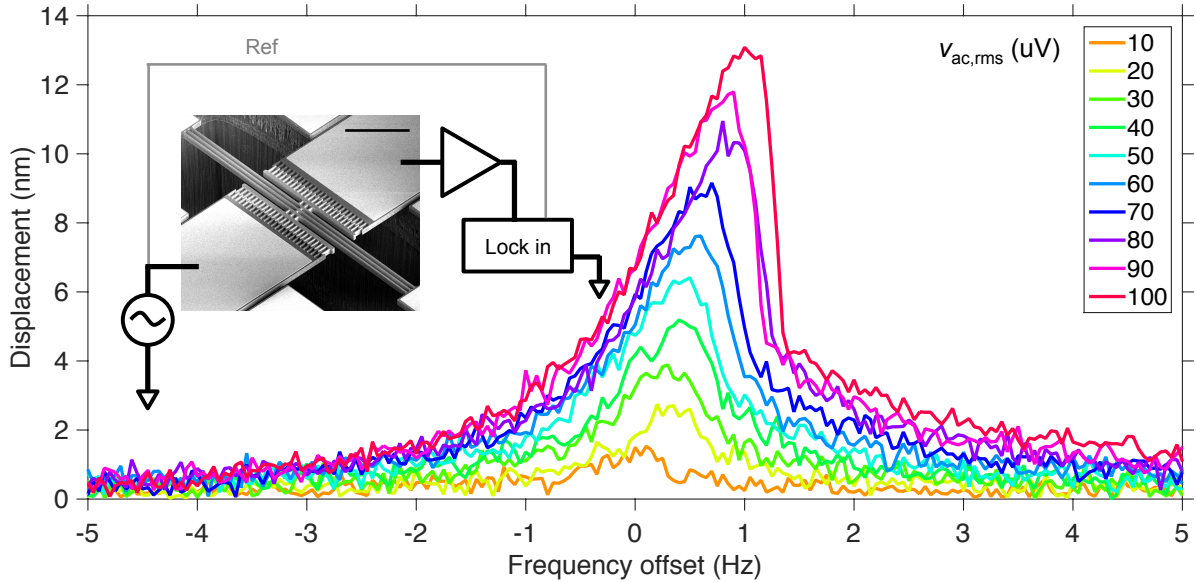

**Figure S6**   Measured mechanical resonance with different ac excitations. The bias voltage is 7 V. Inset: schematic of the open-loop measurement, with the Scanning electron microscopic image of the c-c beam. Scale bar: 100 $\mu$m.

## 5   Experimental evidence of nonlinear damping

The presence of nonlinear damping is critical in the linear velocity feedback scheme. Its existence can be found both from open-loop, resonator-like measurements (quasi-static), and ring-down measurements (dynamical), as described in Section 2. The result is shown in fig. S7 (different device from the one shown in the main text, but with same design). Fig. S7a shows the velocity gain (measured velocity normalized by excitation level) with different excitation level. It is clearly that the maximum velocity gain decreases with increasing excitation, which is the signature of nonlinear damping[2]. We note that the resonator is extremely sensitive to noise around bifurcation, therefore, the velocity may drop from the upper branch at a smaller frequency than the noiseless case. Nevertheless, the observed apparent decrease in maximum velocity gain still reveals the existence of nonlinear damping, and agrees well with other experimental investigation of nonlinear damping in MEMS and NEMS[5,6].

The presence of nonlinear damping is further confirmed by the ring-down measurement, as shown in fig. S7b. During ring-down, the envelop of the vibrational velocity deviates from pure exponential decay that is expected with linear damping, and shows faster-than-exponential decrease during the beginning portion. Such observation agrees well with the theoretical prediction of Eq. (S13) for nonlinear damping, and similar ring-down behavior in silicon MEMS has been reported as well[4]. We have observed similar behavior from all three devices that we tested.

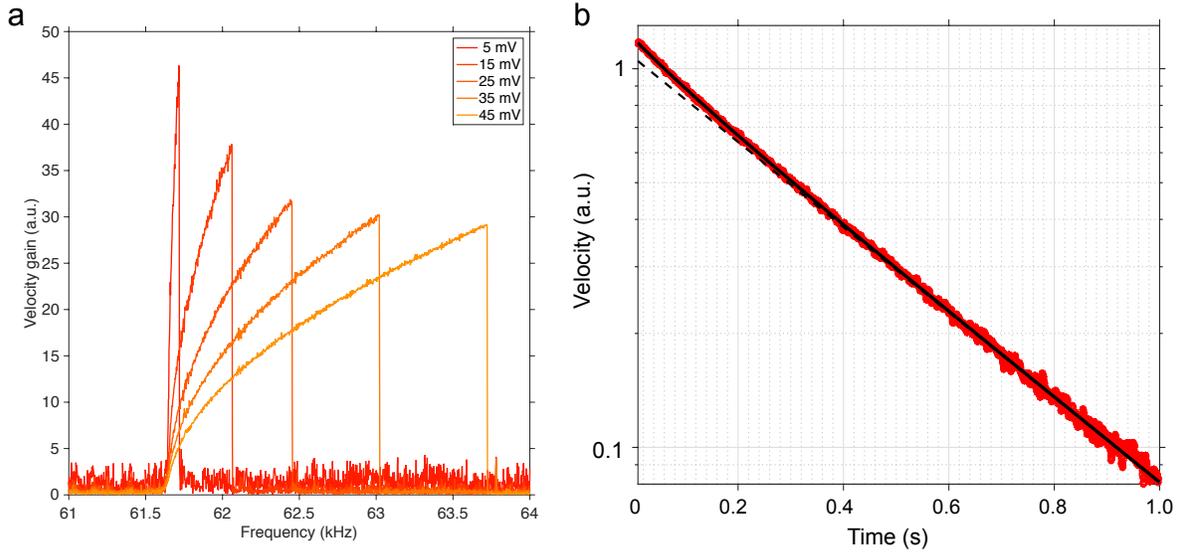

**Figure S7**  Evidences of nonlinear damping. (a) Velocity gain measured with different $v_{ac}$, showing reduced maximum gain. (b) Envelop of the measured velocity during ring-down, with fitting to Eq. (S13) (solid line), with $\eta$ and $q$ as only fitting parameters. The fitted nonlinear damping coefficient is $\tilde{\eta} = (233 \pm 7)$ kg/m$^2$s, and fitted quality factor is $(74.6 \pm 0.1) \times 10^3$. The dashed straight line shows the pure exponential decay with the same quality factor. This quality factor is slightly smaller than the value of $(90.9 \pm 2) \times 10^3$, that is obtained from quasi-static frequency sweep measurement for this device.

## 6  Calibration of feedback circuitry linearities

The feedback circuitry consists of two amplifiers (FEMTO DLPCA-200 current amplifier and Stanford Research System SIM911 voltage amplifier), two analog filters (Stanford Research Sys-

10

tem SIM965), one power splitter (Mini-Circuits ZSC-3-2+), and one RF switch (Mini-Circuits ZASWA-2-50DR+). The analog filters form an effective band-pass filter, with one acts as a low-pass filter and the other acts as a high-pass filter. Additionally, the corner frequency of the high-pass filter is tuned in order to introduce the phase-delay in the feedback loop. The input-output amplitude linearity is tested separately for each component, with a carrier frequency of 65 kHz. When the self-sustained oscillator is in operation, the signals at various locations of the feedback loop are monitored to trace the amplitude. In our experiments, the root mean square (rms) amplitude of the voltage feeding to the voltage amplifier is in the vicinity of 10 mV, which is still well below the onset of nonlinearities of both the preceding transimpedance amplifier (above 200 mV, refer to output), and the following amplifier (about 50 mV, refer to input), as shown in fig. S8.

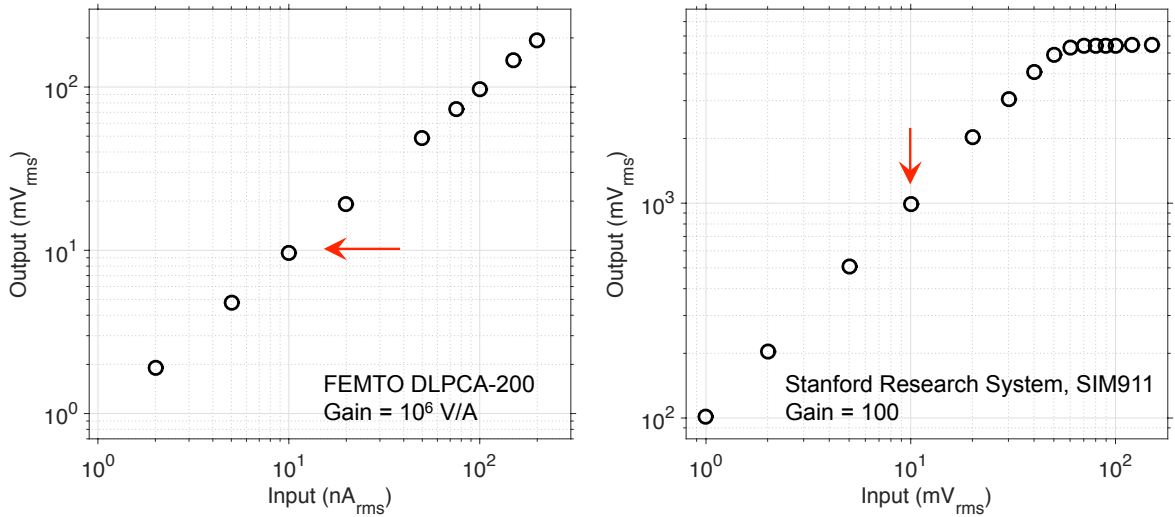

**Figure S8** Input-output linearity calibrations for the amplifiers used in the feedback circuit, the frequency of the input signal is 65 kHz. The red arrows indicate the vicinity of operation ranges in practice.

## 7 Variation of quality factor at different pressures

We varied the linear damping of the constituent resonator, as in fig. 4 of the main text, by changing the pressure of the vacuum chamber. The quality factors, which directly quantified the linear damping rate, are obtained from quasi-static open-loop measurements in the linear regime, at each

11

pressure (fig. S9).

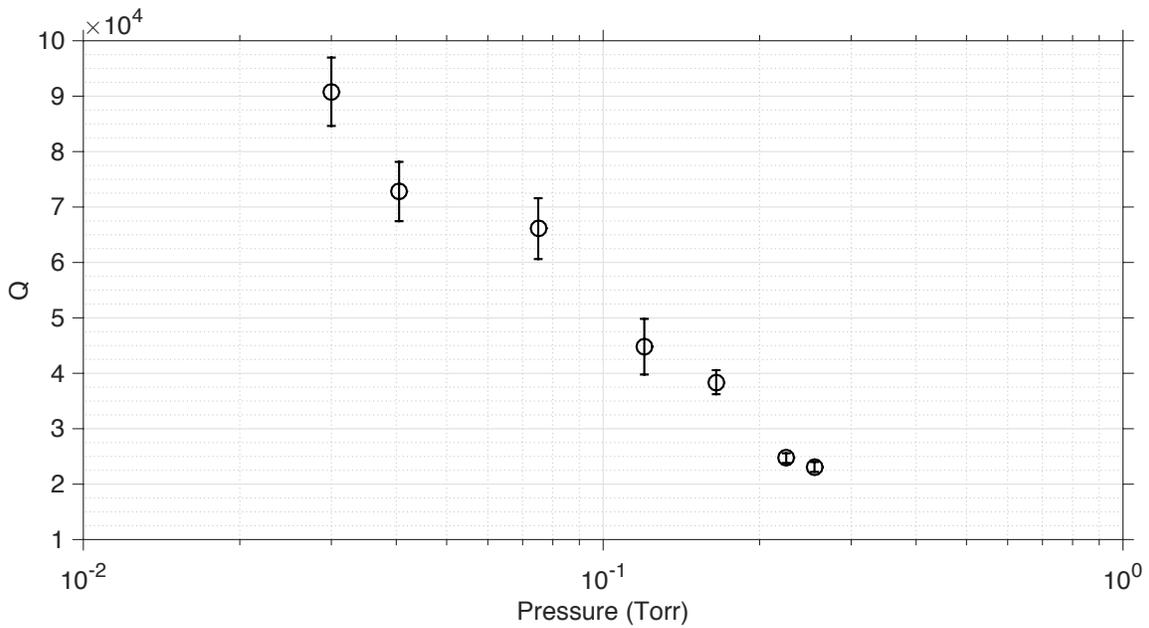

**Figure S9** Quality factors at different pressures.

## 8 Startup time for different velocity gain

The mechanical motion of the c-c beam is actuated electrostatically, where the capacitive force is proportional to the applied DC bias $V_b$. Therefore, in the linear feedback schemes, we can vary the gain by changing $V_b$. Accordingly, the startup time decreases as $V_b$ increases, as shown in fig. 4 in the main text, as well as in fig. S10.

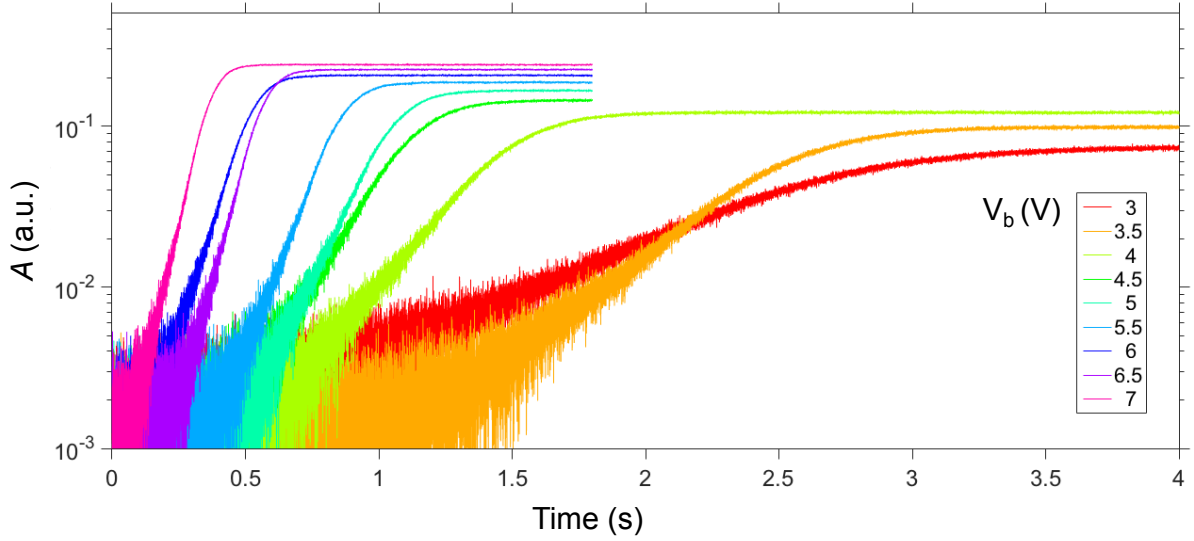

**Figure S10**  Startup of the oscillation with different DC bias voltage $V_b$. The test pressure is below 1 mTorr.

## 9  Startup time versus steady state frequency

The startup time, $t_{\text{startup}}$, is arbitrarily defined as the time needed to reach 90 % of steady state velocity. For the linear velocity feedback case, Eq. (S3) becomes:

$$2U' = gU \sin \Delta - q^{-1} U (1 + \eta A^2), \tag{S14}$$

where $U = A\Omega$ is the velocity. For small $A$, where $\Omega$ can be approximated as 1, we have $U \approx A$, therefore Eq. (S14) becomes:

$$U' = \frac{1}{2} U \left( g \sin \Delta - q^{-1} - \eta q^{-1} U^2 \right). \tag{S15}$$

The solution to Eq. (S15) is:

$$U(T) = \sqrt{\frac{a}{b}} \frac{e^{aT}}{\sqrt{C_1 + e^{2aT}}}, \tag{S16}$$

where $a = \frac{1}{2}(g \sin \Delta - q^{-1})$, $b = \frac{1}{2}\eta q^{-1}$, and $C_1$ is a constant determined by initial conditions. From Eq. S16, we have $U(0) = \sqrt{a/b}\sqrt{C_1 + 1}$, and $U(\infty) = \sqrt{a/b}$. The scaled startup time,

13

$T_\text{startup}$, obtained by solving $U(T_\text{startup}) = 0.9\sqrt{a/b}$, is:

$$\begin{align} T_\text{startup} &= \frac{1}{g\sin\Delta - q^{-1}} \ln(4.26 C_1) \\ &= \frac{\beta}{\eta q^{-1}(\Omega^2 - 1)} \ln(4.26 C_1) \\ &\approx \frac{\beta}{2\eta q^{-1}(\delta\Omega)} \ln(4.26 C_1), \end{align} \tag{S17}$$

where $\delta\Omega = \Omega - 1$. The unscaled startup time is then:

$$t_\text{startup} = \frac{3\epsilon\tilde{\beta}}{2\omega_0 \tilde{\eta}(\delta\omega)} \ln(4.26 C_2), \tag{S18}$$

which is inversely proportional to the frequency offset $\delta\omega$, and independent of feedback gain, feedback phase delay, and linear quality factor.

## 10 Comparison between PLL and linear velocity feedback schemes

To further validate the linear feedback scheme, we also constructed an oscillator with the phase-locked loop (PLL) implementation[7], where in the latter case, the control parameter is the amplitude of the PLL output ($v_{ac}$). We have observed qualitative similar frequency tuning, as shown in fig. S11, however, in order to initiate the oscillation with PLL setup, we need to perform open-loop frequency sweep to bring the system to the appropriate operating point to satisfy Barkhausen criterion, before switching it to the closed-loop configuration.

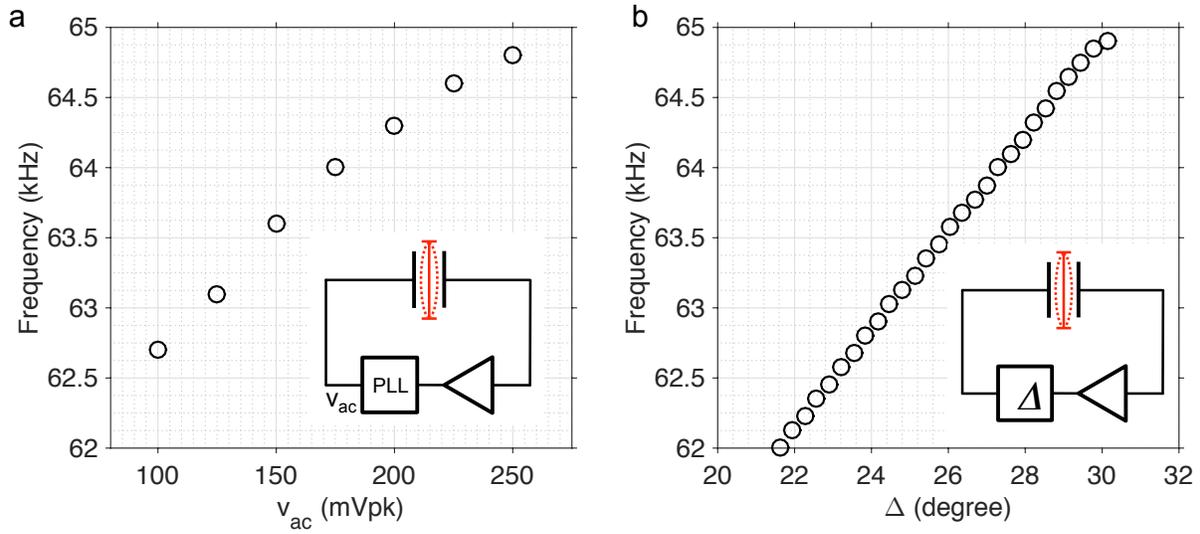

**Figure S11**  Comparison between PLL (a) and linear velocity feedback (b) schemes.

## 11  Calibration of voltage responsivity of in-plane motion

The measured voltage, $V$, is converted to displacement, $A$, through $V = 0.7315 \times V_{DC} \times f \times A$, where $V_{DC}$ is the DC bias, $f$ is the oscillation frequency (in unit of Hz), and the pre-factor is calculated based on the device geometry and the total amplifier gain. Independently, the voltage responsivity of the in-plane motion is calibrated through direct imaging while the c-c beam is oscillating. The vibration amplitude is measured from the gray values of the cross sectional profile (ImageJ). Fig. S12 shows the measured mean vibration with different measured voltage. The length per pixel is calibrated with an additional image while the c-c beam is stationary. This calibrated amplitude agrees well with the calculated values, which are used throughout.

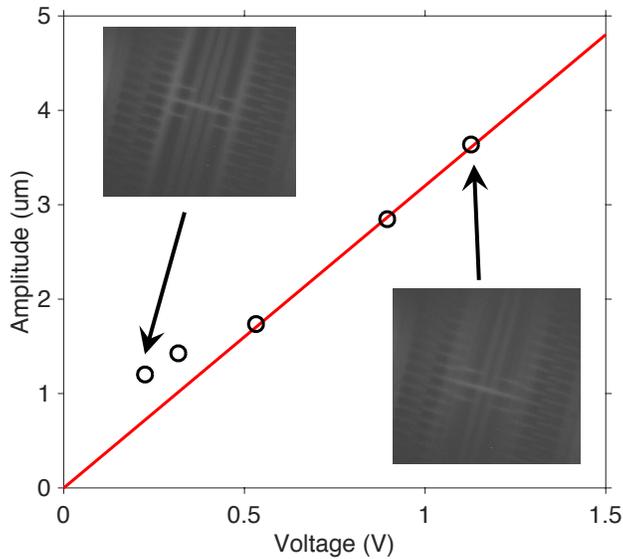

**Figure S12** Calibrated (open circles) and calculated vibrational amplitude. The DC bias is 7 V in both cases, and $f$ is assumed to be 61 kHz in calculation.

1. Kovacic, I. & Brennan, M. J. *The Duffing equation: nonlinear oscillators and their behaviour* (John Wiley & Sons, 2011).

2. Lifshitz, R. & Cross, M. C. Nonlinear dynamics of nanomechanical and micromechanical resonators. In *Reviews of Nonlinear Dynamics and Complexity*, 1–52 (Wiley-VCH Verlag GmbH & Co. KGaA, 2009).

3. Villanueva, L. G. *et al.* A nanoscale parametric feedback oscillator. *Nano Lett.* **11**, 5054–5059 (2011).

4. Polunin, P. M., Yang, Y., Dykman, M. I., Kenny, T. W. & Shaw, S. W. Characterization of mems resonator nonlinearities using the ringdown response. *Journal of Microelectromechanical Systems* **PP**, 1–7 (2016).

5. Imboden, M., Williams, O. A. & Mohanty, P. Observation of nonlinear dissipation in piezoresistive diamond nanomechanical resonators by heterodyne down-mixing. *Nano letters* **13**, 4014–4019 (2013).

6. Imboden, M. & Mohanty, P. Dissipation in nanoelectromechanical systems. *Physics Reports* **534**, 89–146 (2014).

7. Antonio, D., Zanette, D. H. & López, D. Frequency stabilization in nonlinear micromechanical oscillators. *Nature communications* **3**, 806 (2012).




\begin{thebibliography}{36}%
\makeatletter
\providecommand \@ifxundefined [1]{%
 \@ifx{#1\undefined}
}%
\providecommand \@ifnum [1]{%
 \ifnum #1\expandafter \@firstoftwo
 \else \expandafter \@secondoftwo
 \fi
}%
\providecommand \@ifx [1]{%
 \ifx #1\expandafter \@firstoftwo
 \else \expandafter \@secondoftwo
 \fi
}%
\providecommand \natexlab [1]{#1}%
\providecommand \enquote  [1]{``#1''}%
\providecommand \bibnamefont  [1]{#1}%
\providecommand \bibfnamefont [1]{#1}%
\providecommand \citenamefont [1]{#1}%
\providecommand \href@noop [0]{\@secondoftwo}%
\providecommand \href [0]{\begingroup \@sanitize@url \@href}%
\providecommand \@href[1]{\@@startlink{#1}\@@href}%
\providecommand \@@href[1]{\endgroup#1\@@endlink}%
\providecommand \@sanitize@url [0]{\catcode `\\12\catcode `\$12\catcode
  `\&12\catcode `\#12\catcode `\^12\catcode `\_12\catcode `\%12\relax}%
\providecommand \@@startlink[1]{}%
\providecommand \@@endlink[0]{}%
\providecommand \url  [0]{\begingroup\@sanitize@url \@url }%
\providecommand \@url [1]{\endgroup\@href {#1}{\urlprefix }}%
\providecommand \urlprefix  [0]{URL }%
\providecommand \Eprint [0]{\href }%
\providecommand \doibase [0]{http://dx.doi.org/}%
\providecommand \selectlanguage [0]{\@gobble}%
\providecommand \bibinfo  [0]{\@secondoftwo}%
\providecommand \bibfield  [0]{\@secondoftwo}%
\providecommand \translation [1]{[#1]}%
\providecommand \BibitemOpen [0]{}%
\providecommand \bibitemStop [0]{}%
\providecommand \bibitemNoStop [0]{.\EOS\space}%
\providecommand \EOS [0]{\spacefactor3000\relax}%
\providecommand \BibitemShut  [1]{\csname bibitem#1\endcsname}%
\let\auto@bib@innerbib\@empty
\bibitem [{\citenamefont {Strogatz}(2014)}]{Strogatz_book_2014}%
  \BibitemOpen
  \bibfield  {author} {\bibinfo {author} {\bibfnamefont {S.~H.}\ \bibnamefont
  {Strogatz}},\ }\href@noop {} {\emph {\bibinfo {title} {Nonlinear dynamics and
  chaos: with applications to physics, biology, chemistry, and engineering}}}\
  (\bibinfo  {publisher} {Westview press},\ \bibinfo {year} {2014})\BibitemShut
  {NoStop}%
\bibitem [{\citenamefont {Jenkins}(2013)}]{Jenkins_pr_2013}%
  \BibitemOpen
  \bibfield  {author} {\bibinfo {author} {\bibfnamefont {A.}~\bibnamefont
  {Jenkins}},\ }\href@noop {} {\bibfield  {journal} {\bibinfo  {journal}
  {Physics Reports}\ }\textbf {\bibinfo {volume} {525}},\ \bibinfo {pages}
  {167} (\bibinfo {year} {2013})}\BibitemShut {NoStop}%
\bibitem [{\citenamefont {Audoin}\ and\ \citenamefont
  {Guinot}(2001)}]{time_book}%
  \BibitemOpen
  \bibfield  {author} {\bibinfo {author} {\bibfnamefont {C.}~\bibnamefont
  {Audoin}}\ and\ \bibinfo {author} {\bibfnamefont {B.}~\bibnamefont
  {Guinot}},\ }\href@noop {} {\emph {\bibinfo {title} {The measurement of time:
  time, frequency and the atomic clock}}}\ (\bibinfo  {publisher} {Cambridge
  University Press},\ \bibinfo {year} {2001})\BibitemShut {NoStop}%
\bibitem [{\citenamefont {Maiman}(1960)}]{Maiman_nature_1960}%
  \BibitemOpen
  \bibfield  {author} {\bibinfo {author} {\bibfnamefont {T.~H.}\ \bibnamefont
  {Maiman}},\ }\href@noop {} {\bibfield  {journal} {\bibinfo  {journal}
  {Nature}\ }\textbf {\bibinfo {volume} {187}},\ \bibinfo {pages} {493}
  (\bibinfo {year} {1960})}\BibitemShut {NoStop}%
\bibitem [{\citenamefont {Andronov}\ \emph {et~al.}(1966)\citenamefont
  {Andronov}, \citenamefont {Vitt},\ and\ \citenamefont
  {Khaikin}}]{Andronov_book_1966}%
  \BibitemOpen
  \bibfield  {author} {\bibinfo {author} {\bibfnamefont {A.}~\bibnamefont
  {Andronov}}, \bibinfo {author} {\bibfnamefont {A.}~\bibnamefont {Vitt}}, \
  and\ \bibinfo {author} {\bibfnamefont {S.}~\bibnamefont {Khaikin}},\
  }\href@noop {} {\emph {\bibinfo {title} {Theory of oscillators}}},\ Dover
  Books on Electrical Engineering\ (\bibinfo  {publisher} {Dover
  Publications},\ \bibinfo {year} {1966})\BibitemShut {NoStop}%
\bibitem [{\citenamefont {Barkhausen}(1963)}]{Barkhausen_book_1963}%
  \BibitemOpen
  \bibfield  {author} {\bibinfo {author} {\bibfnamefont {H.}~\bibnamefont
  {Barkhausen}},\ }\href@noop {} {\emph {\bibinfo {title} {Lehrbuch der
  Elektronenr{\"o}hren und ihrer technischen Anwendungen}}}\ (\bibinfo
  {publisher} {Hirzel},\ \bibinfo {year} {1963})\BibitemShut {NoStop}%
\bibitem [{\citenamefont {Naik}\ \emph {et~al.}(2009)\citenamefont {Naik},
  \citenamefont {Hanay}, \citenamefont {Hiebert}, \citenamefont {Feng},\ and\
  \citenamefont {Roukes}}]{naik_nnano_2009}%
  \BibitemOpen
  \bibfield  {author} {\bibinfo {author} {\bibfnamefont {A.~K.}\ \bibnamefont
  {Naik}}, \bibinfo {author} {\bibfnamefont {M.~S.}\ \bibnamefont {Hanay}},
  \bibinfo {author} {\bibfnamefont {W.~K.}\ \bibnamefont {Hiebert}}, \bibinfo
  {author} {\bibfnamefont {X.~L.}\ \bibnamefont {Feng}}, \ and\ \bibinfo
  {author} {\bibfnamefont {M.~L.}\ \bibnamefont {Roukes}},\ }\href@noop {}
  {\bibfield  {journal} {\bibinfo  {journal} {Nature Nanotechnology}\ }\textbf
  {\bibinfo {volume} {4}},\ \bibinfo {pages} {445} (\bibinfo {year}
  {2009})}\BibitemShut {NoStop}%
\bibitem [{\citenamefont {Ekinci}\ and\ \citenamefont
  {Roukes}(2005)}]{Ekinci_rsi_2005}%
  \BibitemOpen
  \bibfield  {author} {\bibinfo {author} {\bibfnamefont {K.~L.}\ \bibnamefont
  {Ekinci}}\ and\ \bibinfo {author} {\bibfnamefont {M.~L.}\ \bibnamefont
  {Roukes}},\ } {\bibfield  {journal}
  {\bibinfo  {journal} {Review of Scientific Instruments}\ }\textbf {\bibinfo
  {volume} {76}},\ \bibinfo {pages} {061101} (\bibinfo {year}
  {2005})}\BibitemShut {NoStop}%
\bibitem [{\citenamefont {Cottone}\ \emph {et~al.}(2009)\citenamefont
  {Cottone}, \citenamefont {Vocca},\ and\ \citenamefont
  {Gammaitoni}}]{Cottone_prl_2009}%
  \BibitemOpen
  \bibfield  {author} {\bibinfo {author} {\bibfnamefont {F.}~\bibnamefont
  {Cottone}}, \bibinfo {author} {\bibfnamefont {H.}~\bibnamefont {Vocca}}, \
  and\ \bibinfo {author} {\bibfnamefont {L.}~\bibnamefont {Gammaitoni}},\
  }\href@noop {} {\bibfield  {journal} {\bibinfo  {journal} {Physical Review
  Letters}\ }\textbf {\bibinfo {volume} {102}},\ \bibinfo {pages} {080601}
  (\bibinfo {year} {2009})}\BibitemShut {NoStop}%
\bibitem [{\citenamefont {Schawlow}\ and\ \citenamefont
  {Townes}(1958)}]{Townes_prl_1958}%
  \BibitemOpen
  \bibfield  {author} {\bibinfo {author} {\bibfnamefont {A.~L.}\ \bibnamefont
  {Schawlow}}\ and\ \bibinfo {author} {\bibfnamefont {C.~H.}\ \bibnamefont
  {Townes}},\ }\href@noop {} {\bibfield  {journal} {\bibinfo  {journal}
  {Physical Review}\ }\textbf {\bibinfo {volume} {112}},\ \bibinfo {pages}
  {1940} (\bibinfo {year} {1958})}\BibitemShut {NoStop}%
\bibitem [{\citenamefont {Vittoz}(2010)}]{Vittoz_book_2010}%
  \BibitemOpen
  \bibfield  {author} {\bibinfo {author} {\bibfnamefont {E.}~\bibnamefont
  {Vittoz}},\ }\href@noop {} {\emph {\bibinfo {title} {Low-power crystal and
  MEMS oscillators: the experience of watch developments}}}\ (\bibinfo
  {publisher} {Springer Science \& Business Media},\ \bibinfo {year}
  {2010})\BibitemShut {NoStop}%
\bibitem [{\citenamefont {Feng}\ \emph {et~al.}(2008)\citenamefont {Feng},
  \citenamefont {White}, \citenamefont {Hajimiri},\ and\ \citenamefont
  {Roukes}}]{Feng_nnano_2008_private}%
  \BibitemOpen
  \bibfield  {author} {\bibinfo {author} {\bibfnamefont {X.}~\bibnamefont
  {Feng}}, \bibinfo {author} {\bibfnamefont {C.}~\bibnamefont {White}},
  \bibinfo {author} {\bibfnamefont {A.}~\bibnamefont {Hajimiri}}, \ and\
  \bibinfo {author} {\bibfnamefont {M.~L.}\ \bibnamefont {Roukes}},\
  }\href@noop {} {\bibfield  {journal} {\bibinfo  {journal} {Nature
  nanotechnology}\ }\textbf {\bibinfo {volume} {3}},\ \bibinfo {pages} {342}
  (\bibinfo {year} {2008})},\ \bibinfo {note} {after private communication with
  the author, one of the sustaining amplifier is saturated.}\BibitemShut
  {Stop}%
\bibitem [{\citenamefont {Lee}\ and\ \citenamefont
  {Nguyen}(2003)}]{Lee_ieee_2003}%
  \BibitemOpen
  \bibfield  {author} {\bibinfo {author} {\bibfnamefont {S.}~\bibnamefont
  {Lee}}\ and\ \bibinfo {author} {\bibfnamefont {C.~T.-C.}\ \bibnamefont
  {Nguyen}},\ }in\ \href@noop {} {\emph {\bibinfo {booktitle} {Frequency
  Control Symposium and PDA Exhibition Jointly with the 17th European Frequency
  and Time Forum, 2003. Proceedings of the 2003 IEEE International}}}\
  (\bibinfo {organization} {IEEE},\ \bibinfo {year} {2003})\ pp.\ \bibinfo
  {pages} {341--349}\BibitemShut {NoStop}%
\bibitem [{\citenamefont {Lin}\ \emph {et~al.}(2004)\citenamefont {Lin},
  \citenamefont {Lee}, \citenamefont {Li}, \citenamefont {Xie}, \citenamefont
  {Ren},\ and\ \citenamefont {Nguyen}}]{Lin_ieee_2004}%
  \BibitemOpen
  \bibfield  {author} {\bibinfo {author} {\bibfnamefont {Y.-W.}\ \bibnamefont
  {Lin}}, \bibinfo {author} {\bibfnamefont {S.}~\bibnamefont {Lee}}, \bibinfo
  {author} {\bibfnamefont {S.-S.}\ \bibnamefont {Li}}, \bibinfo {author}
  {\bibfnamefont {Y.}~\bibnamefont {Xie}}, \bibinfo {author} {\bibfnamefont
  {Z.}~\bibnamefont {Ren}}, \ and\ \bibinfo {author} {\bibfnamefont {C.-C.}\
  \bibnamefont {Nguyen}},\ }\href@noop {} {\bibfield  {journal} {\bibinfo
  {journal} {Solid-State Circuits, IEEE Journal of}\ }\textbf {\bibinfo
  {volume} {39}},\ \bibinfo {pages} {2477} (\bibinfo {year}
  {2004})}\BibitemShut {NoStop}%
\bibitem [{\citenamefont {Giessibl}(2003)}]{Giessibl_rmp_2003}%
  \BibitemOpen
  \bibfield  {author} {\bibinfo {author} {\bibfnamefont {F.~J.}\ \bibnamefont
  {Giessibl}},\ }\href@noop {} {\bibfield  {journal} {\bibinfo  {journal}
  {Reviews of modern physics}\ }\textbf {\bibinfo {volume} {75}},\ \bibinfo
  {pages} {949} (\bibinfo {year} {2003})}\BibitemShut {NoStop}%
\bibitem [{\citenamefont {Antonio}\ \emph {et~al.}(2012)\citenamefont
  {Antonio}, \citenamefont {Zanette},\ and\ \citenamefont
  {L{\'o}pez}}]{Antonio_ncomm_2012}%
  \BibitemOpen
  \bibfield  {author} {\bibinfo {author} {\bibfnamefont {D.}~\bibnamefont
  {Antonio}}, \bibinfo {author} {\bibfnamefont {D.~H.}\ \bibnamefont
  {Zanette}}, \ and\ \bibinfo {author} {\bibfnamefont {D.}~\bibnamefont
  {L{\'o}pez}},\ }\href@noop {} {\bibfield  {journal} {\bibinfo  {journal}
  {Nature communications}\ }\textbf {\bibinfo {volume} {3}},\ \bibinfo {pages}
  {806} (\bibinfo {year} {2012})}\BibitemShut {NoStop}%
\bibitem [{\citenamefont {Nguyen}\ and\ \citenamefont
  {Howe}(1999)}]{Nguyen_jssc_1999}%
  \BibitemOpen
  \bibfield  {author} {\bibinfo {author} {\bibfnamefont {C.~T.-C.}\
  \bibnamefont {Nguyen}}\ and\ \bibinfo {author} {\bibfnamefont {R.~T.}\
  \bibnamefont {Howe}},\ }\href@noop {} {\bibfield  {journal} {\bibinfo
  {journal} {Solid-State Circuits, IEEE Journal of}\ }\textbf {\bibinfo
  {volume} {34}},\ \bibinfo {pages} {440} (\bibinfo {year} {1999})}\BibitemShut
  {NoStop}%
\bibitem [{\citenamefont {Villanueva}\ \emph {et~al.}(2011)\citenamefont
  {Villanueva}, \citenamefont {Karabalin}, \citenamefont {Matheny},
  \citenamefont {Kenig}, \citenamefont {Cross},\ and\ \citenamefont
  {Roukes}}]{Villanueva_nl_2011}%
  \BibitemOpen
  \bibfield  {author} {\bibinfo {author} {\bibfnamefont {L.~G.}\ \bibnamefont
  {Villanueva}}, \bibinfo {author} {\bibfnamefont {R.~B.}\ \bibnamefont
  {Karabalin}}, \bibinfo {author} {\bibfnamefont {M.~H.}\ \bibnamefont
  {Matheny}}, \bibinfo {author} {\bibfnamefont {E.}~\bibnamefont {Kenig}},
  \bibinfo {author} {\bibfnamefont {M.~C.}\ \bibnamefont {Cross}}, \ and\
  \bibinfo {author} {\bibfnamefont {M.~L.}\ \bibnamefont {Roukes}},\
  }\href@noop {} {\bibfield  {journal} {\bibinfo  {journal} {Nano Lett.}\
  }\textbf {\bibinfo {volume} {11}},\ \bibinfo {pages} {5054} (\bibinfo {year}
  {2011})}\BibitemShut {NoStop}%
\bibitem [{\citenamefont {Van~Beek}\ and\ \citenamefont
  {Puers}(2011)}]{vanBeek_jmm_2011}%
  \BibitemOpen
  \bibfield  {author} {\bibinfo {author} {\bibfnamefont {J.}~\bibnamefont
  {Van~Beek}}\ and\ \bibinfo {author} {\bibfnamefont {R.}~\bibnamefont
  {Puers}},\ }\href@noop {} {\bibfield  {journal} {\bibinfo  {journal} {Journal
  of Micromechanics and Microengineering}\ }\textbf {\bibinfo {volume} {22}},\
  \bibinfo {pages} {013001} (\bibinfo {year} {2011})}\BibitemShut {NoStop}%
\bibitem [{\citenamefont {Lifshitz}\ and\ \citenamefont
  {Cross}(2009)}]{Lifshitz_2009}%
  \BibitemOpen
  \bibfield  {author} {\bibinfo {author} {\bibfnamefont {R.}~\bibnamefont
  {Lifshitz}}\ and\ \bibinfo {author} {\bibfnamefont {M.~C.}\ \bibnamefont
  {Cross}},\ }in\ \href@noop {} {\emph {\bibinfo {booktitle} {Reviews of
  Nonlinear Dynamics and Complexity}}}\ (\bibinfo  {publisher} {Wiley-VCH
  Verlag GmbH \& Co. KGaA},\ \bibinfo {year} {2009})\ pp.\ \bibinfo {pages}
  {1--52}\BibitemShut {NoStop}%
\bibitem [{\citenamefont {Dykman}\ and\ \citenamefont
  {Krivoglaz}(1975)}]{Dykman_pss_1975}%
  \BibitemOpen
  \bibfield  {author} {\bibinfo {author} {\bibfnamefont {M.}~\bibnamefont
  {Dykman}}\ and\ \bibinfo {author} {\bibfnamefont {M.}~\bibnamefont
  {Krivoglaz}},\ }\href@noop {} {\bibfield  {journal} {\bibinfo  {journal}
  {physica status solidi (b)}\ }\textbf {\bibinfo {volume} {68}},\ \bibinfo
  {pages} {111} (\bibinfo {year} {1975})}\BibitemShut {NoStop}%
\bibitem [{\citenamefont {Dykman}(2012)}]{Dykman_book_2012}%
  \BibitemOpen
  \bibfield  {author} {\bibinfo {author} {\bibfnamefont {M.}~\bibnamefont
  {Dykman}},\ }\href@noop {} {\emph {\bibinfo {title} {Fluctuating nonlinear
  oscillators: from nanomechanics to quantum superconducting circuits}}}\
  (\bibinfo  {publisher} {OUP Oxford},\ \bibinfo {year} {2012})\BibitemShut
  {NoStop}%
\bibitem [{prl()}]{prl_SI}%
  \BibitemOpen
  \href@noop {} {\bibinfo  {journal} {See Supplementary Information, which
  includes Refs. [35-36]}\ }\BibitemShut {NoStop}%
\bibitem [{\citenamefont {Nayfeh}\ and\ \citenamefont
  {Mook}(2008)}]{Nayfeh_book_2008}%
  \BibitemOpen
\bibfield  {journal} {  }\bibfield  {author} {\bibinfo {author} {\bibfnamefont
  {A.~H.}\ \bibnamefont {Nayfeh}}\ and\ \bibinfo {author} {\bibfnamefont
  {D.~T.}\ \bibnamefont {Mook}},\ }\href@noop {} {\emph {\bibinfo {title}
  {Nonlinear oscillations}}}\ (\bibinfo  {publisher} {John Wiley \& Sons},\
  \bibinfo {year} {2008})\BibitemShut {NoStop}%
\bibitem [{\citenamefont {Arroyo}\ and\ \citenamefont
  {Zanette}(2016)}]{Arroyo_epjb_2016}%
  \BibitemOpen
  \bibfield  {author} {\bibinfo {author} {\bibfnamefont {S.~I.}\ \bibnamefont
  {Arroyo}}\ and\ \bibinfo {author} {\bibfnamefont {D.~H.}\ \bibnamefont
  {Zanette}},\ }\href@noop {} {\bibfield  {journal} {\bibinfo  {journal} {The
  European Physical Journal B}\ }\textbf {\bibinfo {volume} {89}},\ \bibinfo
  {pages} {1} (\bibinfo {year} {2016})}\BibitemShut {NoStop}%
\bibitem [{\citenamefont {Villanueva}\ \emph {et~al.}(2013)\citenamefont
  {Villanueva}, \citenamefont {Kenig}, \citenamefont {Karabalin}, \citenamefont
  {Matheny}, \citenamefont {Lifshitz}, \citenamefont {Cross},\ and\
  \citenamefont {Roukes}}]{Villanueva_prl_2013}%
  \BibitemOpen
  \bibfield  {author} {\bibinfo {author} {\bibfnamefont {L.~G.}\ \bibnamefont
  {Villanueva}}, \bibinfo {author} {\bibfnamefont {E.}~\bibnamefont {Kenig}},
  \bibinfo {author} {\bibfnamefont {R.~B.}\ \bibnamefont {Karabalin}}, \bibinfo
  {author} {\bibfnamefont {M.~H.}\ \bibnamefont {Matheny}}, \bibinfo {author}
  {\bibfnamefont {R.}~\bibnamefont {Lifshitz}}, \bibinfo {author}
  {\bibfnamefont {M.~C.}\ \bibnamefont {Cross}}, \ and\ \bibinfo {author}
  {\bibfnamefont {M.~L.}\ \bibnamefont {Roukes}},\ }\href@noop {} {\bibfield
  {journal} {\bibinfo  {journal} {Phys. Rev. Lett.}\ }\textbf {\bibinfo
  {volume} {110}},\ \bibinfo {pages} {177208} (\bibinfo {year}
  {2013})}\BibitemShut {NoStop}%
\bibitem [{\citenamefont {Imboden}\ and\ \citenamefont
  {Mohanty}(2014)}]{Imboden_pr_2014}%
  \BibitemOpen
  \bibfield  {author} {\bibinfo {author} {\bibfnamefont {M.}~\bibnamefont
  {Imboden}}\ and\ \bibinfo {author} {\bibfnamefont {P.}~\bibnamefont
  {Mohanty}},\ }\href@noop {} {\bibfield  {journal} {\bibinfo  {journal}
  {Physics Reports}\ }\textbf {\bibinfo {volume} {534}},\ \bibinfo {pages} {89}
  (\bibinfo {year} {2014})}\BibitemShut {NoStop}%
\bibitem [{\citenamefont {Polunin}\ \emph {et~al.}(2016)\citenamefont
  {Polunin}, \citenamefont {Yang}, \citenamefont {Dykman}, \citenamefont
  {Kenny},\ and\ \citenamefont {Shaw}}]{Polunin_jmems_2016}%
  \BibitemOpen
  \bibfield  {author} {\bibinfo {author} {\bibfnamefont {P.~M.}\ \bibnamefont
  {Polunin}}, \bibinfo {author} {\bibfnamefont {Y.}~\bibnamefont {Yang}},
  \bibinfo {author} {\bibfnamefont {M.~I.}\ \bibnamefont {Dykman}}, \bibinfo
  {author} {\bibfnamefont {T.~W.}\ \bibnamefont {Kenny}}, \ and\ \bibinfo
  {author} {\bibfnamefont {S.~W.}\ \bibnamefont {Shaw}},\ } {\bibfield  {journal} {\bibinfo  {journal}
  {Journal of Microelectromechanical Systems}\ }\textbf {\bibinfo {volume}
  {PP}},\ \bibinfo {pages} {1} (\bibinfo {year} {2016})}\BibitemShut {NoStop}%
\bibitem [{\citenamefont {Sedra}\ and\ \citenamefont
  {Smith}(1998)}]{Sedra_book_1998}%
  \BibitemOpen
  \bibfield  {author} {\bibinfo {author} {\bibfnamefont {A.~S.}\ \bibnamefont
  {Sedra}}\ and\ \bibinfo {author} {\bibfnamefont {K.~C.}\ \bibnamefont
  {Smith}},\ }\href@noop {} {\emph {\bibinfo {title} {Microelectronic
  circuits}}},\ Vol.~\bibinfo {volume} {1}\ (\bibinfo  {publisher} {New York:
  Oxford University Press},\ \bibinfo {year} {1998})\BibitemShut {NoStop}%
\bibitem [{\citenamefont {Nguyen}(2007)}]{Nguyen_ieee_2007}%
  \BibitemOpen
  \bibfield  {author} {\bibinfo {author} {\bibfnamefont {C.-C.}\ \bibnamefont
  {Nguyen}},\ }\href@noop {} {\bibfield  {journal} {\bibinfo  {journal}
  {Ultrasonics, Ferroelectrics and Frequency Control, IEEE Transactions on}\
  }\textbf {\bibinfo {volume} {54}},\ \bibinfo {pages} {251} (\bibinfo {year}
  {2007})}\BibitemShut {NoStop}%
\bibitem [{\citenamefont {Antonio}\ \emph {et~al.}(2015)\citenamefont
  {Antonio}, \citenamefont {Czaplewski}, \citenamefont {Guest}, \citenamefont
  {L{\'o}pez}, \citenamefont {Arroyo},\ and\ \citenamefont
  {Zanette}}]{Antonio_prl_2015}%
  \BibitemOpen
  \bibfield  {author} {\bibinfo {author} {\bibfnamefont {D.}~\bibnamefont
  {Antonio}}, \bibinfo {author} {\bibfnamefont {D.~A.}\ \bibnamefont
  {Czaplewski}}, \bibinfo {author} {\bibfnamefont {J.~R.}\ \bibnamefont
  {Guest}}, \bibinfo {author} {\bibfnamefont {D.}~\bibnamefont {L{\'o}pez}},
  \bibinfo {author} {\bibfnamefont {S.~I.}\ \bibnamefont {Arroyo}}, \ and\
  \bibinfo {author} {\bibfnamefont {D.~H.}\ \bibnamefont {Zanette}},\
  }\href@noop {} {\bibfield  {journal} {\bibinfo  {journal} {Physical review
  letters}\ }\textbf {\bibinfo {volume} {114}},\ \bibinfo {pages} {034103}
  (\bibinfo {year} {2015})}\BibitemShut {NoStop}%
\bibitem [{\citenamefont {Gieseler}\ \emph {et~al.}(2013)\citenamefont
  {Gieseler}, \citenamefont {Novotny},\ and\ \citenamefont
  {Quidant}}]{Gieseler_nphys_2013}%
  \BibitemOpen
  \bibfield  {author} {\bibinfo {author} {\bibfnamefont {J.}~\bibnamefont
  {Gieseler}}, \bibinfo {author} {\bibfnamefont {L.}~\bibnamefont {Novotny}}, \
  and\ \bibinfo {author} {\bibfnamefont {R.}~\bibnamefont {Quidant}},\
  }\href@noop {} {\bibfield  {journal} {\bibinfo  {journal} {Nature Physics}\
  }\textbf {\bibinfo {volume} {9}},\ \bibinfo {pages} {806} (\bibinfo {year}
  {2013})}\BibitemShut {NoStop}%
\bibitem [{\citenamefont {Postma}\ \emph {et~al.}(2005)\citenamefont {Postma},
  \citenamefont {Kozinsky}, \citenamefont {Husain},\ and\ \citenamefont
  {Roukes}}]{Postma_apl_2005}%
  \BibitemOpen
  \bibfield  {author} {\bibinfo {author} {\bibfnamefont {H.~C.}\ \bibnamefont
  {Postma}}, \bibinfo {author} {\bibfnamefont {I.}~\bibnamefont {Kozinsky}},
  \bibinfo {author} {\bibfnamefont {A.}~\bibnamefont {Husain}}, \ and\ \bibinfo
  {author} {\bibfnamefont {M.}~\bibnamefont {Roukes}},\ }\href@noop {}
  {\bibfield  {journal} {\bibinfo  {journal} {Applied Physics Letters}\
  }\textbf {\bibinfo {volume} {86}},\ \bibinfo {pages} {223105} (\bibinfo
  {year} {2005})}\BibitemShut {NoStop}%
\bibitem [{\citenamefont {Chen}\ \emph {et~al.}(2013)\citenamefont {Chen},
  \citenamefont {Lee}, \citenamefont {Deshpande}, \citenamefont {Lee},
  \citenamefont {Lekas}, \citenamefont {Shepard},\ and\ \citenamefont
  {Hone}}]{Chen_nnano_2013}%
  \BibitemOpen
  \bibfield  {author} {\bibinfo {author} {\bibfnamefont {C.}~\bibnamefont
  {Chen}}, \bibinfo {author} {\bibfnamefont {S.}~\bibnamefont {Lee}}, \bibinfo
  {author} {\bibfnamefont {V.~V.}\ \bibnamefont {Deshpande}}, \bibinfo {author}
  {\bibfnamefont {G.-H.}\ \bibnamefont {Lee}}, \bibinfo {author} {\bibfnamefont
  {M.}~\bibnamefont {Lekas}}, \bibinfo {author} {\bibfnamefont
  {K.}~\bibnamefont {Shepard}}, \ and\ \bibinfo {author} {\bibfnamefont
  {J.}~\bibnamefont {Hone}},\ }\href@noop {} {\bibfield  {journal} {\bibinfo
  {journal} {Nature Nano}\ }\textbf {\bibinfo {volume} {8}},\ \bibinfo {pages}
  {923} (\bibinfo {year} {2013})}\BibitemShut {NoStop}%
\bibitem [{\citenamefont {Kovacic}\ and\ \citenamefont
  {Brennan}(2011)}]{Kovacic_book_2011}%
  \BibitemOpen
  \bibfield  {author} {\bibinfo {author} {\bibfnamefont {I.}~\bibnamefont
  {Kovacic}}\ and\ \bibinfo {author} {\bibfnamefont {M.~J.}\ \bibnamefont
  {Brennan}},\ }\href@noop {} {\emph {\bibinfo {title} {The Duffing equation:
  nonlinear oscillators and their behaviour}}}\ (\bibinfo  {publisher} {John
  Wiley \& Sons},\ \bibinfo {year} {2011})\BibitemShut {NoStop}%
\bibitem [{\citenamefont {Imboden}\ \emph {et~al.}(2013)\citenamefont
  {Imboden}, \citenamefont {Williams},\ and\ \citenamefont
  {Mohanty}}]{Imboden_nl_2013}%
  \BibitemOpen
  \bibfield  {author} {\bibinfo {author} {\bibfnamefont {M.}~\bibnamefont
  {Imboden}}, \bibinfo {author} {\bibfnamefont {O.~A.}\ \bibnamefont
  {Williams}}, \ and\ \bibinfo {author} {\bibfnamefont {P.}~\bibnamefont
  {Mohanty}},\ }\href@noop {} {\bibfield  {journal} {\bibinfo  {journal} {Nano
  letters}\ }\textbf {\bibinfo {volume} {13}},\ \bibinfo {pages} {4014}
  (\bibinfo {year} {2013})}\BibitemShut {NoStop}%
\end{thebibliography}
\end{document}